\documentclass[journal,11pt,draftclsnofoot,onecolumn]{IEEEtran}
\usepackage{cite}

\ifCLASSINFOpdf
\else
\fi
\usepackage[pdftex]{graphicx}

\usepackage{amsmath}

\renewcommand{\IEEEQED}{\IEEEQEDopen}

\usepackage{amssymb}
\usepackage{color}
\newtheorem{definition}{Definition}[section]
\newtheorem{proposition}{Proposition}[section] 
\newtheorem{theorem}{Theorem}[section]
\newtheorem{corollary}{Corollary}[section]
\newtheorem{lemma}{Lemma}[section]

\newtheorem{remark}{Remark}[section]

\begin{document}
%
\title{Optimum Achievable Rates in Two Random Number Generation Problems with $f$-Divergences Using Smooth R\'enyi Entropy}
%
%
%

\author{Ryo Nomura,~\IEEEmembership{Member,~IEEE}  
        and~Hideki Yagi,~\IEEEmembership{Member,~IEEE}
\thanks{R. Nomura is with the Center for Data Science, Waseda University, Tokyo 169-8050, Japan, e-mail: nomu@waseda.jp}
\thanks{H. Yagi is with the Department of Computer and Network Engineering,  
the University of Electro-Communications, Tokyo 184-8795, Japan, 
e-mail: h.yagi@uec.ac.jp}
\thanks{This paper is an extended version of the conference papers \cite{NY_ISIT2020} and \cite{NY_ISIT2021}.}
}
\maketitle

\begin{abstract}
Two typical fixed-length random number generation problems in information theory are considered for \textit{general} sources. One is the source resolvability problem and the other is the intrinsic randomness problem.
In each of these problems, the optimum achievable rate with respect to the given approximation measure is one of our main concerns and has been characterized using two different information quantities: the information spectrum and the smooth R\'enyi entropy.
Recently, optimum achievable rates with respect to $f$-divergences have been characterized using the information spectrum quantity.
The $f$-divergence is a general non-negative measure between two probability distributions on the basis of a convex function $f$. The class of $f$-divergences includes several important measures such as the variational distance, the KL divergence, the Hellinger distance and so on. 
Hence, it is meaningful to consider the random number generation problems with respect to $f$-divergences.
However, optimum achievable rates with respect to $f$-divergences using the smooth R\'enyi entropy have not been clarified yet in both of two problems.
In this paper we try to analyze the optimum achievable rates using the smooth R\'enyi entropy and to extend the class of $f$-divergence.
To do so, we first derive general formulas of the \textit{first-order} optimum achievable rates with respect to $f$-divergences in both problems under the same conditions as imposed by previous studies.
Next, we relax the conditions on $f$-divergence and generalize the obtained general formulas.
Then, we particularize our general formulas to several specified functions $f$. 
As a result, we reveal that it is easy to derive optimum achievable rates for several important measures from our general formulas. Furthermore, a kind of \textit{duality} between the resolvability and the intrinsic randomness is revealed in terms of the smooth R\'enyi entropy.
\textit{Second-order} optimum achievable rates and optimistic achievable rates are also investigated.
\end{abstract}


%

\section{Introduction}
%
%
%
%
\IEEEPARstart{T}wo typical fixed-length random number generation problems in information theory are considered for \textit{general} sources. One is the source resolvability problem (the resolvability problem for short) and the other is the intrinsic randomness problem.
The problem setting of the {\it resolvability} problem is as follows.
Given an arbitrary source ${\bf X} = \{ X^n \}_{n=1}^\infty$ (the {\it target} random number), we approximate it by using a discrete random number which is uniformly distributed, which we call the \textit{uniform random number}. Here, the size of the uniform random number is requested to be as small as possible.
In this setting, a degree of approximation is measured by several criteria. 
Han and Verd\'{u} \cite{HV93}, and Steinberg and Verd\'{u} \cite{Steinberg} have determined the \textit{first-order} optimum achievable rates with respect to the variational distance and the {normalized} Kullback-Leibler (KL) divergence. 
Nomura \cite{Nomura_ISIT2018} has studied the \textit{first-order} optimum achievable rates with respect to the KL divergence.
Recently, Nomura \cite{Nomura_TIT2020} has characterized the \textit{first-order} optimum achievable rates with respect to $f$-divergences.
The class of $f$-divergence considered in \cite{Nomura_TIT2020} includes the variational distance and the KL divergence. Hence, the result can be considered as a generalization of results given in \cite{HV93} and \cite{Nomura_ISIT2018}.
The \textit{second-order} optimum achievable rates in the resolvability problem have also been studied with respect to several approximation measures \cite{NH2011,Nomura_TIT2020}. 
It should be noted that  results mentioned above are based on the information spectrum quantity. 
On the other hand, Uyematsu \cite{Uyematsu_ISIT2010} has characterized the \textit{first-order} optimum achievable rate with respect to the variational distance using the smooth R\'enyi entropy.

The \textit{intrinsic randomness} problem, which is also one of typical random number generation problems, has also been studied.
The problem setting of the intrinsic randomness problem is as follows.
By using a given arbitrary source ${\bf X} = \{ X^n \}_{n=1}^\infty$ (the {\it coin} random number), we approximate a discrete {\it uniform} random number whose size is requested to be as large as possible.
Also in the intrinsic randomness problem, optimum achievable rates with respect to various criteria have been considered.
Vembu and Verd\'{u} \cite{VV} have considered the intrinsic randomness problem with respect to the variational distance as well as the normalized KL divergence and derived \textit{general formulas} of the \textit{first-order} optimum achievable rates (cf. Han \cite{Han}).
Hayashi \cite{Hayashi} has considered the \textit{first-} and \textit{second-order} optimum achievable rates with respect to the KL divergence.
Recently, the \textit{first-} and \textit{second-order}  optimum achievable rates with respect to $f$-divergences have been clarified in \cite{Nomura_TIT2020}.
The results mentioned here are based on information spectrum quantities.
On the other hand, Uyematsu and Kunimatsu \cite{Uyematsu13} have characterized the \textit{first-order} optimum achievable rates with respect to the variational distance using the smooth R\'enyi entropy.

Related works include works given by Liu, Cuff and Verd\'u \cite{LCV2017}, Yagi and Han \cite{YH_arxiv2018}, Kumagai and Hayashi \cite{Kumagai2017a, Kumagai2017b}, and Yu and Tan \cite{YT18}.
In \cite{LCV2017}, the \textit{channel} resolvability problem with respect to the $E_\gamma$-divergence has been considered.
They have applied their results to the case of the \textit{source} resolvability problem.
Yagi and Han \cite{YH2023} have determined the optimum \textit{variable-length} resolvability rates with respect to the variational distance as well as the KL divergence.  
Kumagai and Hayashi \cite{Kumagai2017a, Kumagai2017b} have determined the \textit{first- and second- order} optimum achievable rates in the \textit{random number conversion} problem.
It should be noted that the random number conversion problem includes the resolvability and intrinsic randomness problems treated in this paper. In \cite{Kumagai2017a} and \cite{Kumagai2017b}, an approximation measure related to the Hellinger distance has been used.
Yu and Tan \cite{YT18} have considered the \textit{random number conversion} problem with respect to the R\'enyi divergence.

As we have mentioned above, in both problem of the resolvability and the intrinsic randomness, various approximation measures have been considered. 
Furthermore, general formulas of achievable rates have been characterized by using the information spectrum quantity and the smooth R\'enyi entropy.
We here note that optimum achievable rates with respect to $f$-divergence using the smooth R\'enyi entropy have not been clarified yet.
The smooth R\'enyi entropy is an information quantity that has a clear operational meaning and is easy to understand.
Moreover, a class of $f$-divergences is a general distance measure, in which several important measures are included.
In this paper, hence, we try to characterize the \textit{first-} and \textit{second-order} optimum achievable rates with respect to $f$-divergences using the smooth R\'enyi entropy.
In addition, we also extend the class of $f$-divergence for which optimum achievable rates can be characterized.
As a result, we find that two types of smooth R\'enyi entropies are useful to describe these optimum achievable rates for a wider class of $f$-divergence.
Furthermore, a kind of \textit{duality} between the resolvability and the intrinsic randomness is revealed in terms of the smooth R\'enyi entropy and $f$-divergences.

This paper is organized as follows.
In Section II, we describe the problem setting and give some definitions of the optimum \textit{first}-order achievable rates. 
The class of $f$-divergences and the smooth R\'enyi entropy have also been introduced.
In Section III and IV, we show \textit{general formulas} of the optimum \textit{first-order} achievable rates in the resolvability problem and the intrinsic randomness problem, respectively.
In Section V, we derive the general formulas of these achievable rates for an extended class of $f$-divergence.
In Section VI, we apply general formulas obtained in previous sections to some specified functions $f$ and compute the optimum \textit{first-order} achievable rates in each cases.
In Section VII, we show \textit{general} formulas of the optimum \textit{second}-order achievable rates in two problems.
In Section  VIII, optimum achievable rates in the optimistic sense are considered.
Section IX is devoted to the discussion concerning our results.
Finally, we provide some concluding remarks on our results in Section X.
\section{Preliminaries}
\subsection{$f$-divergences}
The $f$-divergence between two probability distributions $P_{Z}$ and $P_{\overline{Z}}$ is defined as follows \cite{csiszar2004information}.
Let $f(t)$ be a convex function defined for $t >0$ and $f(1) =0$.
\begin{definition}[$f$-divergence \cite{csiszar2004information}]  \label{def:f-divergence}
Let $P_Z$ and $P_{\overline{Z}}$ denote probability distributions over a finite or countably infinite set ${\cal Z}$. The $f$-divergence between $P_{Z}$ and $P_{\overline{Z}}$ is defined by
\begin{equation}
D_f(Z||\overline{Z}) :=  \sum_{z \in {{\cal Z}}} P_{\overline{Z}}(z) f \left(\frac{P_{{Z}}(z)}{P_{\overline{Z}}(z)}\right),
\end{equation}
where we set $0 f\left(\frac{0}{0}\right) =0$,  $f(0) = \lim_{t \downarrow 0} f(t)$, $0f(\frac{a}{0}) = \lim_{t \to 0} t f(\frac{a}{t}) = a \lim_{u \to \infty}\frac{f(u)}{u}$.
\end{definition}
The $f$-divergence is a general approximation measure, which includes some important measures.
We give some examples of $f$-divergences \cite{csiszar2004information,SV2016}:
\begin{itemize}
\item $f(t) = t \log t$:  (Kullback-Leibler (KL) divergence)
\begin{IEEEeqnarray}{rCl}
D_f(Z||\overline{Z}) =   \sum_{z \in {{\cal Z}}} P_{{Z}}(z) \log \frac{P_{{Z}}(z)}{P_{\overline{Z}}(z)} =: D(Z||\overline{Z}).
\end{IEEEeqnarray}
\item $f(t) = - \log t$:  (Reverse Kullback-Leibler divergence)
\begin{IEEEeqnarray}{rCl}
D_f(Z||\overline{Z}) =   \sum_{z \in {{\cal Z}}} P_{\overline{Z}}(z) \log \frac{P_{\overline{Z}}(z)}{P_{{Z}}(z)} = D(\overline{Z}||{Z}).
\end{IEEEeqnarray}
\item $f(t) = 1 - \sqrt{t}$:  (Hellinger distance)
\begin{IEEEeqnarray}{rCl}
D_f(Z||\overline{Z}) =   1 - \sum_{z \in {{\cal Z}}}\sqrt{P_{{Z}}(z) P_{\overline{Z}}(z)}.
\end{IEEEeqnarray}
\item $f(t) = (1- \sqrt{t})^2$: (Squared Hellinger distance)
\begin{IEEEeqnarray}{rCl}
D_f(Z||\overline{Z}) =   \sum_{z \in {{\cal Z}}} \left( \sqrt{P_{{Z}}(z)}  - \sqrt{P_{\overline{Z}}(z)} \right)^2.
\end{IEEEeqnarray}
\item $f(t) = |t-1|$:  (Variational distance)
\begin{IEEEeqnarray}{rCl}
D_f(Z||\overline{Z}) =   \sum_{z \in {{\cal Z}}}  |P_{{Z}}(z) - P_{\overline{Z}}(z)|.
\end{IEEEeqnarray}
\item $f(t) = (1-t)^+ := \max\{1-t,0\}$:  (Half variational distance)
\begin{IEEEeqnarray}{rCl}  \label{eq:hvd}
D_f(Z||\overline{Z}) & = & \frac{1}{2} \sum_{z \in {{\cal Z}}}  |P_{{Z}}(z) - P_{\overline{Z}}(z)| \nonumber \\
& = & \sum_{z \in {{\cal Z}}: P_{{Z}}(z) > P_{\overline{Z}}(z)} \left(P_{{Z}}(z) - P_{\overline{Z}}(z)\right).
\end{IEEEeqnarray}
\item $f(t) = \frac{t^\alpha - \alpha t -(1-\alpha)}{\alpha(\alpha-1)}$: $\alpha$-divergence ($0 < \alpha <1$)
\begin{IEEEeqnarray}{rCl} 
D_f(Z||\overline{Z}) & = & \frac{1}{\alpha(\alpha-1)}\left( 1 - \sum_{z \in {\cal Z}} P_{{Z}}(z)^{\alpha} P_{\overline{Z}}(z)^{1-\alpha} \right).
\end{IEEEeqnarray}
\item $f(t) = (t-\gamma)^+$ :  ($E_\gamma$-divergence)
For any given $\gamma \ge1$, 
\begin{IEEEeqnarray}{rCl} \label{eq:egamma}
D_f(Z||\overline{Z}) =  \sum_{z \in {{\cal Z}}: P_{{Z}}(z) > \gamma P_{\overline{Z}}(z)} \left(P_{{Z}}(z) - \gamma P_{\overline{Z}}(z)\right) =: E_\gamma(Z||\overline{Z}) .
\end{IEEEeqnarray}
\end{itemize}
The $E_\gamma$-divergence is a generalization of the half variational distance defined in (\ref{eq:hvd}), because $\gamma \ge 1$ is arbitrarily.
\begin{remark}
It is known \cite{Nomura_TIT2020} that the $E_{\gamma}$-divergence can be expressed as an $f$-divergence using the function:
\begin{equation} \label{eq:egamma2}
f(t) = (\gamma - t)^+ + 1 -\gamma.
\end{equation}
\end{remark}

The following key property holds for the $f$-divergence from Jensen's inequality \cite{csiszar2004information}:
\begin{IEEEeqnarray}{rCl} \label{eq:log-sum}
\sum_{z \in {\cal Z}'} b(z) f\left(\frac{a(z)}{b(z)} \right) 
 & \ge & \left(\sum_{z \in {\cal Z}'} b(z) \right)  f\left(\frac{\sum_{z \in {\cal Z}'}a(z)}{\sum_{z \in {\cal Z}'} b(z)} \right).
\end{IEEEeqnarray}

As we have mentioned above, the $f$-divergence is a general approximation measure, which includes several important measures.
In this study, we {first} assume the following conditions on the function $f$, which are also imposed by previous studies \cite{Nomura_TIT2020}.
\begin{description}
\item[C1)] The function $f(t)$ is a decreasing function for $t >0$ with $f(0) >0$. 
\item[C2)] The function $f(t)$ satisfies 
\begin{equation} \label{assumption:4}
\lim_{u \to \infty} \frac{f(u)}{u} =0.
\end{equation}
\item[C3)] For any pair of positive real numbers $(a,b)$, it holds that
\begin{equation} \label{assumption:2}
\lim_{n\to \infty} \frac{f\left(e^{-nb} \right)}{e^{na}} = 0.
\end{equation}
\end{description}
\begin{remark}  \label{remark:conditions}
Notice here that functions $f(t) = -\log t$, $f(t) = 1 - \sqrt{t}$, and $f(t) = ( 1- t)^+$ satisfy the above conditions, while $f(t) = t \log t$ does not satisfy conditions C1) and C2).
Moreover, it is not difficult to check that (\ref{eq:egamma2}) satisfies these conditions.
\end{remark}
\begin{remark} 
For a decreasing function $f(t)$, it always holds that $f(0) = \lim_{t \downarrow 0} f(t) \ge 0$ because $f(1) = 0$. Then, the condtion $f(0) > 0$ in C1) excludes the case of $f(t) = 0$ for all $t \ge 0$, in which $f$-divergence is identically zero. 
\end{remark}


\begin{remark}  \label{remark:assump}
From the definition of the $f$-divergence, C2) means
\begin{equation} \label{assumption:5}
0 f\left(\frac{a}{0}\right) = 0,
\end{equation}
for any $a > 0$.
In the derivation of our main theorems, we can use (\ref{assumption:5}) instead of (\ref{assumption:4}).
\end{remark}
\begin{remark} \label{remark:relaxed_condition}
We will show in Section \ref{sect:relaxed_condition} that condition C1) is automatically met for the function $f$ satisfying condition C2) (cf.\ claim (i) of Lemma \ref{lem:offset_f}).
\end{remark}
\subsection{Smooth {R}\'enyi entropy}
In what follows, we consider the case of ${\cal Z} = {\cal X}^n$, where ${\cal X}$ is a finite or countably infinite set and $n$ is an integer.
We consider the \textit{general} source defined as an infinite sequence 
\[
{\bf X} = \left\{X^n = \left(X_1^{(n)}, X_2^{(n)}, \dots, X_n^{(n)}\right) \right\}_{n=1}^\infty
\] of $n$-dimensional random variables $X^n$, where each component random variable $X^{(n)}_i$ takes values in a countable set ${\cal X}$. 
Let $P_X(\cdot)$ denote the probability distribution of the random variable $X$.
In this paper, we assume the following condition on the source ${\bf X}$:
\begin{equation} \label{eq:assump_source}
\underline{H}({\bf X}) < +\infty,
\end{equation}
where 
\begin{IEEEeqnarray}{rCl}
\underline{H}({\bf X}) 
&= & \sup\left\{ R \left| \lim_{n \to \infty} \Pr\left\{\frac{1}{n} \log \frac{1}{P_{X^n}(X^n)} \ge R \right\} = 1   \right. \right\}
\end{IEEEeqnarray}
is called the spectral inf-entropy rate of the source ${\bf X}$ \cite{Han}.
Here, Han \cite[Theorem 1.7.2]{Han} has shown that 
\begin{IEEEeqnarray}{rCl}
\underline{H}({\bf X}) \le \log |{\cal X}|
\end{IEEEeqnarray}
holds. 
Hence, the condition (\ref{eq:assump_source}) holds for any source with \textit{finite alphabet}.

The random number $U_{M}$ which is uniformly distributed on $\{1,2,\cdots, M \}$ is defined by
\begin{equation}
P_{U_{M}}(i) = \frac{1}{M}, \ \ i \in {\cal U}_M := \{1,2,\cdots, M \}.
\end{equation}

We next introduce the smooth R\'enyi entropy of the source.
\begin{definition}[Smooth R\'enyi entropy of order $\alpha$ \cite{RW2004}]
For given random variables $X^n$, the smooth R\'enyi entropy of order $\alpha$ given $\delta \ (0 \le \delta <1)$ is defined by
\begin{equation} \label{eq:renyi}
H_{\alpha}(\delta| {X^n}) := \frac{1}{1 - \alpha} \inf_{ P_{\overline{X}^n} \in B^\delta(P_{X^n})} \log \left( \sum_{{\bf x} \in {\cal X}^n} P_{\overline{X}^n}({\bf x})^\alpha  \right),
\end{equation}
where
\begin{IEEEeqnarray}{rCl} 
B^\delta(P_{X^n}) 
& := & \left\{ P_{\overline{X}^n} \in {\cal P}^n \left| \frac{1}{2} \sum_{{\bf x} \in {\cal X}^n} |P_{X^n}({\bf x}) - P_{\overline{X}^n}({\bf x})  |  \le \delta \right. \right\}.\ \ 
\end{IEEEeqnarray}
\end{definition}
Here, $H_{\alpha}(\delta| {X^n})$ is a decreasing function of $\delta$.
The smooth  R\'enyi entropy of order $0$ and  the smooth  R\'enyi entropy of order $\infty$ are respectively called the smooth max entropy and the smooth min entropy \cite{HR2011}.

The following theorems have shown alternative expressions of the smooth  max entropy and the smooth  min entropy.
\begin{theorem}[Uyematsu \cite{Uyematsu_ISIT2010,Uyematsu2010} ]  \label{theo:Uye}
\begin{equation}  \label{eq:SRE0}
H_{0}(\delta| {X^n}) = \min_{\substack{A_n \subset {\cal X}^n \\ \Pr\{X^n \in A_n \} \ge 1 - \delta  }} \log |A_n|.
\end{equation}
\end{theorem}
\begin{theorem}[Uyematsu and Kunimatsu \cite{Uyematsu13}]  \label{theo:Uye2}
\begin{equation}  \label{eq:SREI}
H_{\infty}(\delta| {X^n}) =  -\inf_{\substack{\beta \ge \frac{1}{|{\cal X}^n|} :  \\ \sum_{{\bf x} \in {\cal X}^n} (P_{X^n}({\bf x}) - \beta)^+    \le  \delta}  } \log {\beta},
\end{equation}
where if $|{\cal X}|$ is a countably infinite set, the infimum is taken over $\beta \ge 0$.
\end{theorem}

\begin{figure}[b]
\begin{center}
\includegraphics[width=2.6in]{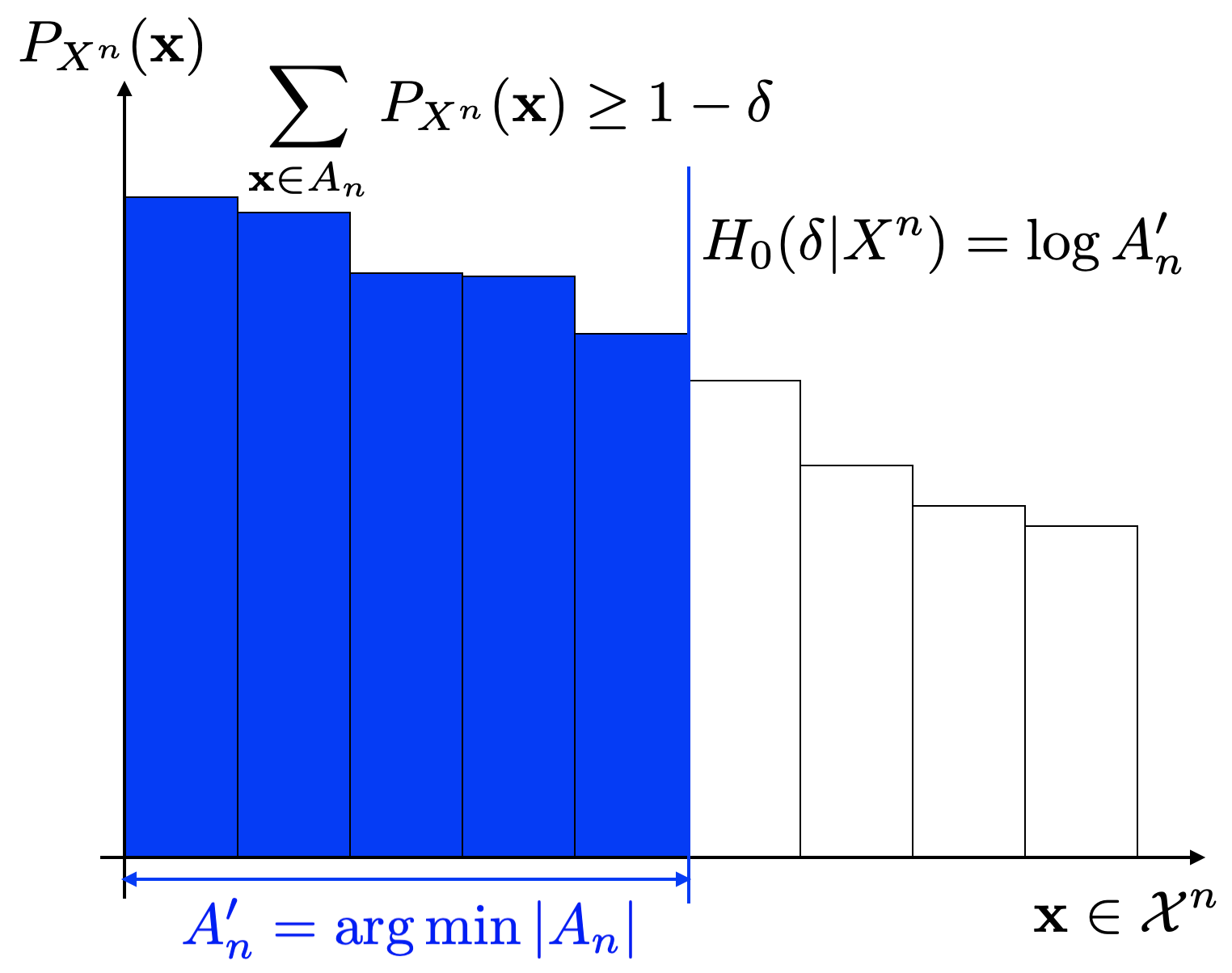}
\centering \caption{Smooth max entropy $H_0(\delta|X^n)$}
\label{fig:0}
\end{center}
\end{figure}

\begin{figure}[t]
\begin{center}
\includegraphics[width=2.6in]{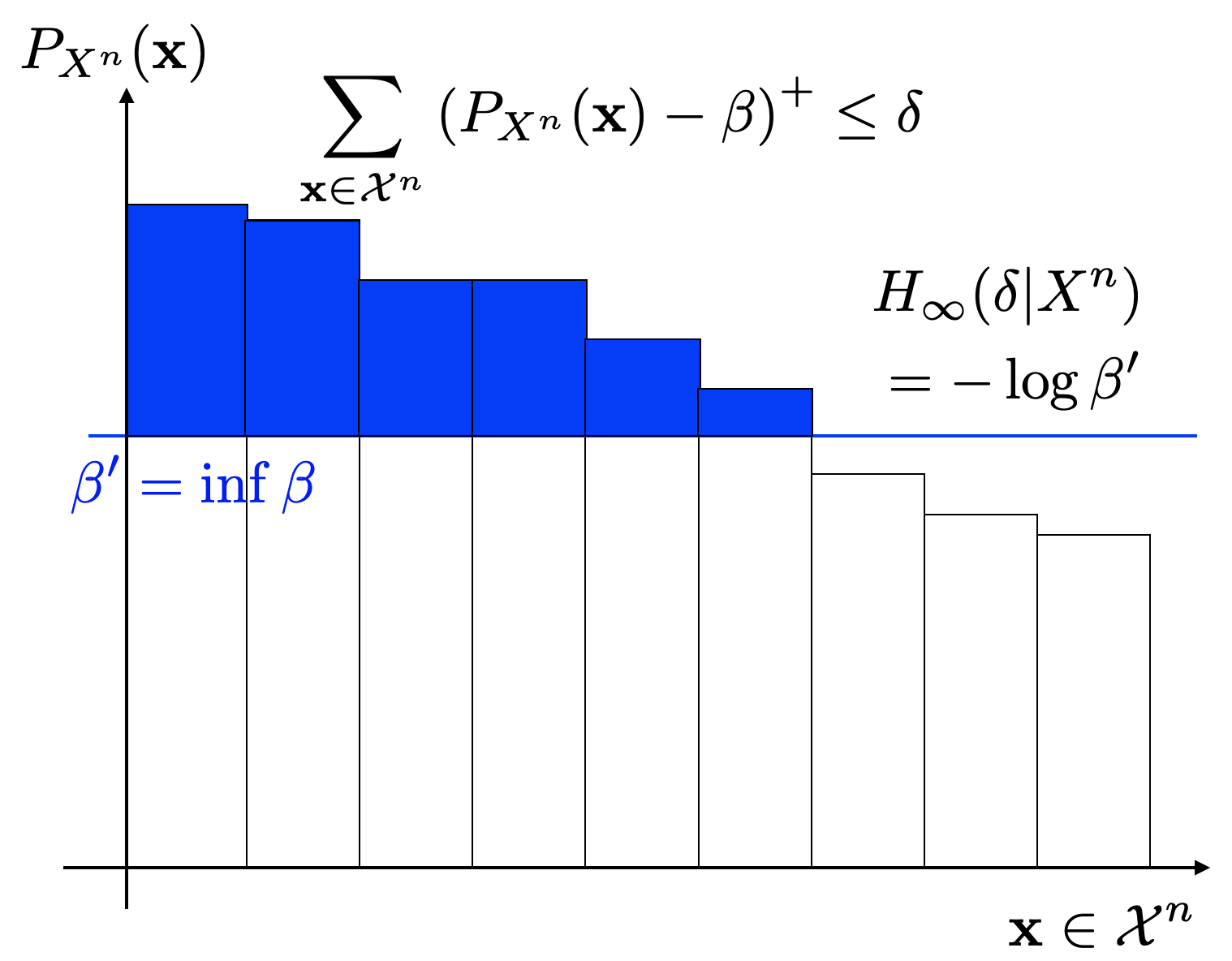}
\centering \caption{Smooth min entropy $H_\infty(\delta|X^n)$}
\label{fig:infty}
\end{center}
\end{figure}

It should be noted that these alternative expressions are simple and easy to understand compared to  (\ref{eq:renyi}).
Fig. \ref{fig:0} and Fig. \ref{fig:infty} show operational meanings of (\ref{eq:SRE0}) and (\ref{eq:SREI}).
As in Fig. \ref{fig:0}, the smooth max entropy $H_{0}(\delta| {X^n})$ is equal to the logarithm of the cardinality of the set $A_n$ with $\Pr\{X^n \in A_n\} \ge 1 - \delta$ where each of the sequence ${\bf x} \in A_n$ has large probability.
On the other hand, the smooth min entropy $H_{\infty}(\delta| {X^n})$ is equal to the supremum of $- \log \beta$ such that the sum of probabilities of sequences ${\bf x} \in {\cal X}^n$ that exceeds $\beta$ is less than or equal to $\delta$ (Fig. \ref{fig:infty}).

In this paper, we use the above alternative expressions of the smooth max entropy and the smooth min entropy,  instead of (\ref{eq:renyi}).

%
%
%
%
%
%
%
\section{Source Resolvability problem}
\begin{figure}[t]
\begin{center}
\includegraphics[width=4.2in]{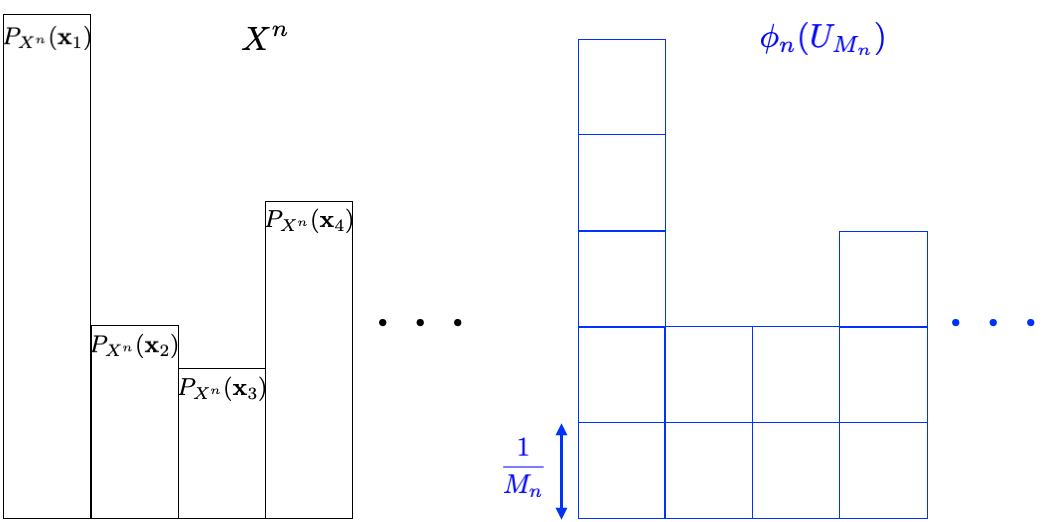}
\centering \caption{Resolvability problem}
\label{fig:SR}
\end{center}
\end{figure}

We consider the problem concerning how to simulate a given discrete source ${\bf X} = \{ X^n \}_{n=1}^\infty$ by using the uniform random number $U_{M_n}$ and the mapping $\phi_n$.
Fig. \ref{fig:SR} is an illustrative figure of this problem.
Since it is hard to simulate the source exactly in general, we consider the approximation problem under some measure.
This problem is called the resolvability problem.
One of main objectives in the resolvability problem is to derive the smallest value of $a$ in the form of $M_n = e^{na}$, which we call the optimum resolvability rate \cite{HV93,Han}.
This is formulated as follows.
\begin{definition} \label{def:3-1} 
Rate $R$ is said to be $D$-achievable with the given $f$-divergence  if there exists a sequence of mapping $\phi_n : {\cal U}_{M_n} \to {\cal X}^n$ such that 
\begin{equation}
\limsup_{n \rightarrow \infty} D_f(X^n||\phi_n(U_{M_n})) \leq D,
\end{equation}
\begin{equation}
\limsup_{n \rightarrow \infty} \frac{1}{n} \log M_n \leq R.
\end{equation}
\end{definition}

Given some $D$, if the rate constraint $R$ is sufficiently large, it can be shown that there exists a sequence of  mappings satisfying constraints in the above definition. Conversely, if $R$ is too small, no sequence of mappings that satisfies constraints can be found. Therefore, in the resolvability problem, the infimum of $R$ is of particular interest.
\begin{definition}[First-order optimum resolvability rate] \label{def:Su}
\begin{IEEEeqnarray}{rCl} 
 S_r^{(f)}(D|{\bf X}) & := & \inf \left\{ R \left|R \mbox{ is $D$-achievable with the given $f$-divergence}  \right. \right\}.
\end{IEEEeqnarray}
\end{definition}
\begin{remark} \label{remark:employ1}
It should be noted that we do not use $D_f(\phi_n(U_{M_n})||X^n)$ but $D_f(X^n||\phi_n(U_{M_n}))$ as a condition in Def. \ref{def:3-1}.
This is important to consider the \textit{asymmetric} measure such as the KL-divergence.
\end{remark}
\begin{remark} \label{remark:trivial}
We consider the case where $D$ is in $[0, f(0))$ under the given $f$-divergence.
Since $f(t)$ is defined in the range $t > 0$ and we assume that the function $f(t)$ is a decreasing function of $t$, $D_f(X^n|| Y^n) \le f(0)$ holds for any distributions $P_{X^n}(\cdot)$ and $P_{Y^n}(\cdot)$ from the definition of $f$-divergence.
 Hence, $D \ge f(0)$ means that there exists no restriction about the approximation error (for example, $f(0)=1$ in the case of the half variational distance and $f(0)=\infty$ in the case of the KL divergence). This case leads to the trivial result that the first-order optimum resolvability rate equals to $0$.
 Hence, we only consider the case of $D \in [0,f(0))$.
The similar observation is applicable throughout the following sections.
\end{remark}

Our main objective in this section is to derive the general formula of the first-order optimum resolvability rate.
To do so, we first derive the following two theorems. 
We use the notation $f^{-1}(a) = \inf\{t | f(t) = a \}.$

\begin{theorem} \label{theo:direct}
Under conditions C1)--C3), for any  $\gamma >0$ and any $M_n$ satisfying
\begin{equation} \label{eq:assump1}
\frac{1}{n} \log M_n \ge \frac{1}{n}H_0(1 - f^{-1}(D)|X^n) + \gamma,
\end{equation}
there exists a mapping $\phi_n$ which satisfies
\begin{equation}  \label{eq:theorem_1-2}
D_f(X^n||\phi_n(U_{M_n})) \le D + \gamma
\end{equation}
for sufficiently large $n$.
\end{theorem}

\begin{IEEEproof}
We arbitrarily fix $M_n$ satisfying (\ref{eq:assump1}). We show that there exists a mapping $\phi_n$ that satisfies (\ref{eq:theorem_1-2}) for sufficiently large $n$.
Let $B_n \subset {\cal X}^n$ denote a set satisfying
\begin{equation} \label{eq:a-0-0}
\Pr\{X^n \in B_n \} \ge  f^{-1}(D)
\end{equation}
and
\begin{equation} \label{eq:b-b-1}
\log |B_n| = H_0(1 - f^{-1}(D)|X^n).
\end{equation}
The existence of the above set $B_n$ is guaranteed by (\ref{eq:SRE0}).
We define the probability distribution $P_{\overline{X}^n}$ over $B_n$ as
\begin{equation}
P_{\overline{X}^n}({\bf x}) := \left\{ \begin{array}{cc}
\frac{P_{{X}^n}({\bf x})}{\Pr\{ X^n \in B_n \}} & {\bf x} \in B_n, \\
0 & \mbox{otherwise}.
\end{array} \right.
\end{equation}
Furthermore, let a set $C_n$ be as
\begin{equation}
C_n := \left\{ {\bf x} \in B_n \left| P_{\overline{X}^n}({\bf x}) \ge \frac{1}{M_n} \right.\right\}
\end{equation}
and arrange elements in $C_n$ as
\begin{equation}
C_n =\{{\bf x}_1,{\bf x}_2, \dots, {\bf x}_{|C_n|} \}
\end{equation}
according to $P_{\overline{X}^n}({\bf x})$ in ascendant order.
That is, $P_{\overline{X}^n}({\bf x}_i) \le P_{\overline{X}^n}({\bf x}_j) \ (1\le i < j\le |C_n|)$ holds.
Here, we define $i^\ast := |C_n|$ and index ${\bf x}  \in B_n \setminus C_n$ as
${\bf x}_{i^\ast + 1}, {\bf x}_{i^\ast + 2},\dots, {\bf x}_{|B_n|}$ arbitrarily.

Then, from the above definition, it holds that
\begin{equation}
P_{\overline{X}^n}({\bf x}_{i^\ast}) = \max_{{\bf x} \in C_n} P_{\overline{X}^n}({\bf x}).
\end{equation}
Thus, from the assumption (\ref{eq:assump_source}), for any small $\varepsilon \in (0,\underline{H}({\bf X}))$, it holds that 
\begin{equation}  \label{eq:assum_source2}
P_{\overline{X}^n}({\bf x}_{i^\ast}) \ge e^{- n (\underline{H}({\bf X}) - \varepsilon)}
\end{equation}
for sufficiently large $n$.

Set $k_0=0$. For ${\bf x}_1$ we determine $k_1$ such that 
\begin{equation}
\frac{k_1}{M_n} \le P_{\overline{X}^n}({\bf x}_1), \quad \frac{k_1+1}{M_n} >  P_{\overline{X}^n}({\bf x}_1).
\end{equation}
Secondly, we determine $k_2$ for $ {\bf x}_2$ such that 
\begin{equation}
\frac{k_2-k_1}{M_n} \le P_{\overline{X}^n}({\bf x}_2), \quad \frac{k_2-k_1+1}{M_n} > P_{\overline{X}^n}({\bf x}_2).
\end{equation}
In the similar way, we repeat this operation to choose $k_i$ for ${\bf x}_i$ as long as possible. 
Then, it is not difficult to check that the above procedure does not stop before $i < i^\ast$.

We define a mapping $\phi_n: {\cal U}_{M_n} \to {\cal X}^n$ as
\begin{equation}
\phi_n(j) = \left\{ \begin{array}{ll}
{\bf x}_i & k_{i-1}+ 1 \le j \le k_i,  \  i < i^\ast \\
{\bf x}_{i^\ast} & \mbox{otherwise}
\end{array} \right.
\end{equation}
and set $\tilde{X}^n = \phi_n(U_{M_n})$.

We evaluate the performance of the mapping $\phi_n$.
From the construction of the mapping, for any $i$ satisfying $1 \le  i\le i^\ast-1$ it holds that
\begin{equation} \label{eq:a-1}
P_{\tilde{X}^n} ({\bf x}_i) \le P_{\overline{X}^n} ({\bf x}_i) 
\end{equation}
\begin{equation}  \label{eq:a-2}
P_{\overline{X}^n} ({\bf x}_i) < P_{\tilde{X}^n} ({\bf x}_i)  +  \frac{1}{M_n}.
\end{equation}
We next evaluate $P_{\tilde{X}^n}({\bf x}_{i^\ast})$. 
From the construction, we have $P_{\overline{X}^n}({\bf x}_{i^\ast}) \le P_{\tilde{X}^n}({\bf x}_{i^\ast})$.
Since
$
P_{\tilde{X}^n}({\bf x}_i) = 0
$
holds for $ \forall i \in B_n \setminus C_n$, we obtain
\begin{equation} \label{eq:a-3-2}
P_{\overline{X}^n}({\bf x}_i) - P_{\tilde{X}^n}({\bf x}_i) = P_{\overline{X}^n}({\bf x}_i)< \frac{1}{M_n}
\end{equation}
for $ \forall i \in B_n \setminus C_n$.
Hence, also from the construction of the mapping, we obtain
\begin{IEEEeqnarray}{rCl}  \label{eq:a-3}
P_{\tilde{X}^n} ({\bf x}_{i^\ast}) - P_{\overline{X}^n} ({\bf x}_{i^\ast})
& = & \left( 1 - \sum_{i=1}^{i^\ast-1} P_{\tilde{X}^n} ({\bf x}_i) \right) - \left( 1 - \sum_{{\bf x}_i \in B_n \setminus\{{\bf x}_{i^\ast}\}  } P_{\overline{X}^n} ({\bf x}_i) \right) \nonumber \\
& =& \sum_{{\bf x}_i \in B_n \setminus\{{\bf x}_{i^\ast}\}  } P_{\overline{X}^n} ({\bf x}_i) - \sum_{{\bf x}_i \in B_n \setminus\{{\bf x}_{i^\ast}\}  } P_{\tilde{X}^n} ({\bf x}_i) \nonumber \\
& = & \sum_{{\bf x}_i \in B_n \setminus\{{\bf x}_{i^\ast}\}  } \left( P_{\overline{X}^n} ({\bf x}_i) -  P_{\tilde{X}^n} ({\bf x}_i) \right) \nonumber \\
& \le &\frac{|B_n|}{M_n} \nonumber \\
& \le & e^{-n \gamma},
\end{IEEEeqnarray}
where the second equality is from the fact that $P_{\tilde{X}^n}({\bf x}_i) = 0$ for  $ \forall i \in B_n \setminus C_n$,  the first inequality is due to (\ref{eq:a-2}) and (\ref{eq:a-3-2}), and the last inequality is obtained from 
(\ref{eq:assump1}) and (\ref{eq:b-b-1}).
Thus, we have
\begin{equation} \label{eq:a-a-4}
P_{\tilde{X}^n} ({\bf x}_{i^\ast}) \le P_{\overline{X}^n} ({\bf x}_{i^\ast}) + e^{-n\gamma}.
\end{equation}

From the above argument the $f$-divergence is given by
\begin{IEEEeqnarray}{rCl}  \label{eq:a-4}
D_f\left( {X^n}||\phi_n(U_{M_n}) \right)
 & = & \sum_{i=1}^{i^\ast} P_{\tilde{X}^n} ({\bf x}_i) f \left(\frac{P_{{X}^n} ({\bf x}_i)}{P_{\tilde{X}^n} ({\bf x}_i)} \right) \nonumber \\
& = & \sum_{i=1}^{i^\ast} P_{\tilde{X}^n} ({\bf x}_i) f\left( \frac{P_{\overline{X}^n} ({\bf x}_i) \Pr\left\{ X^n \in B_n\right\}}{P_{\tilde{X}^n} ({\bf x}_i) } \right) \nonumber \\
& = & \sum_{i=1}^{i^\ast -1} P_{\tilde{X}^n} ({\bf x}_i) f\left( \frac{P_{\overline{X}^n} ({\bf x}_i) \Pr\left\{ X^n \in B_n\right\}}{P_{\tilde{X}^n} ({\bf x}_i) } \right) \nonumber \\
&  & + P_{\tilde{X}^n} ({\bf x}_{i^\ast}) f\left( \frac{P_{\overline{X}^n} ({\bf x}_{i^\ast}) \Pr\left\{ X^n \in B_n\right\}}{P_{\tilde{X}^n} ({\bf x}_{i^\ast}) } \right)\nonumber \\
& \le & \sum_{i=1}^{i^\ast -1} P_{\tilde{X}^n} ({\bf x}_i) f\left( \Pr\left\{ X^n \in B_n\right\} \right)  \nonumber \\
&&  + P_{\tilde{X}^n} ({\bf x}_{i^\ast}) f\left( \frac{P_{\overline{X}^n} ({\bf x}_{i^\ast}) \Pr\left\{ X^n \in B_n\right\}}{P_{\tilde{X}^n} ({\bf x}_{i^\ast}) } \right),  
\end{IEEEeqnarray}
where the first equality is due to the condition C2) and the last inequality is due to (\ref{eq:a-1}) and the condition C1).

The second term of the RHS of (\ref{eq:a-4}) is evaluated as follows.
From (\ref{eq:a-a-4}) and C1) we have
\begin{IEEEeqnarray}{rCl} \label{eq:39}
\lefteqn{P_{\tilde{X}^n} ({\bf x}_{i^\ast}) f\left( \frac{P_{\overline{X}^n} ({\bf x}_{i^\ast}) \Pr\left\{ X^n \in B_n\right\}}{P_{\tilde{X}^n} ({\bf x}_{i^\ast}) } \right)} \nonumber  \\
& \le & \left(P_{\overline{X}^n} ({\bf x}_{i^\ast}) + e^{-n \gamma}\right) f\left( \frac{P_{\overline{X}^n} ({\bf x}_{i^\ast}) \Pr\left\{ X^n \in B_n\right\}}{P_{\overline{X}^n} ({\bf x}_{i^\ast}) + e^{-n\gamma} } \right).
\end{IEEEeqnarray}

Here, using the relation 
\begin{IEEEeqnarray}{rCl} \label{eq:f12}
\lefteqn{P_{\overline{X}^n} ({\bf x}_{i^\ast}) \Pr\left\{ X^n \in B_n\right\} } \nonumber \\
& = & (1-e^{-n\gamma})P_{\overline{X}^n} ({\bf x}_{i^\ast}) \Pr\left\{ X^n \in B_n\right\}  +  e^{-n \gamma}P_{\overline{X}^n} ({\bf x}_{i^\ast}) \Pr\left\{ X^n \in B_n\right\},
 \end{IEEEeqnarray}
we obtain
\begin{IEEEeqnarray}{rCl}
\lefteqn{\left(P_{\overline{X}^n} ({\bf x}_{i^\ast}) + e^{-n\gamma}\right) f\left( \frac{P_{\overline{X}^n} ({\bf x}_{i^\ast}) \Pr\left\{ X^n \in B_n\right\}}{P_{\overline{X}^n} ({\bf x}_{i^\ast}) + e^{-n\gamma} } \right)} \nonumber \\
 & \le & P_{\overline{X}^n} ({\bf x}_{i^\ast})  f\left( \frac{ (1 - e^{-n\gamma} )P_{\overline{X}^n} ({\bf x}_{i^\ast}) \Pr\left\{ X^n \in B_n\right\}}{P_{\overline{X}^n} ({\bf x}_{i^\ast}) } \right)  \nonumber \\
 & & + e^{-n\gamma} f\left( \frac{ e^{-n\gamma} P_{\overline{X}^n} ({\bf x}_{i^\ast}) \Pr\left\{ X^n \in B_n\right\}}{e^{-n\gamma} } \right) \nonumber \\
& = & P_{\overline{X}^n} ({\bf x}_{i^\ast})  f\left( (1 - e^{-n\gamma} ) \Pr\left\{ X^n \in B_n\right\} \right) \nonumber \\
& & + e^{-n\gamma} f\left( P_{\overline{X}^n} ({\bf x}_{i^\ast}) \Pr\left\{ X^n \in B_n\right\} \right) \nonumber \\
& \le & P_{\overline{X}^n} ({\bf x}_{i^\ast})  f\left( (1 - e^{-n\gamma} ) \Pr\left\{ X^n \in B_n\right\} \right)  \nonumber \\
& & + e^{-n\gamma} f\left( e^{-n(\underline{H}({\bf X}) - \varepsilon) } \Pr\left\{ X^n \in B_n\right\} \right) 
\end{IEEEeqnarray}
for sufficiently large $n$, where the first inequality is due to (\ref{eq:log-sum}) and the last inequality is from 
(\ref{eq:assum_source2}) and the condition C1).

Hence, from C3) and the continuity of the function $f$, for $\forall \nu >0$ we have 
\begin{IEEEeqnarray}{rCl} \label{eq:a-4-2}
\lefteqn{\left(P_{\overline{X}^n} ({\bf x}_{i^\ast}) + e^{-n\gamma}\right) f\left( \frac{P_{\overline{X}^n} ({\bf x}_{i^\ast}) \Pr\left\{ X^n \in B_n\right\}}{P_{\overline{X}^n} ({\bf x}_{i^\ast}) + e^{-n\gamma} } \right)} \nonumber \\
& \le & P_{\overline{X}^n} ({\bf x}_{i^\ast})  f\left(  \Pr\left\{ X^n \in B_n\right\}  - e^{-n\gamma} \right)   + \nu \nonumber \\
& \le &  P_{\overline{X}^n} ({\bf x}_{i^\ast})  f\left(  \Pr\left\{ X^n \in B_n\right\}\right)   + 2\nu
\end{IEEEeqnarray}
for sufficiently large $n$.
Therefore, noting that $P_{\overline{X}^n} ({\bf x}_{i^\ast})  \le P_{\tilde{X}^n} ({\bf x}_{i^\ast})$, from (\ref{eq:a-0-0}), (\ref{eq:a-4}), (\ref{eq:39}) and (\ref{eq:a-4-2}) it holds that
\begin{IEEEeqnarray}{rCl}  \label{eq:a-6}
D_f\left( {X^n}||\phi_n(U_{M_n}) \right)
 & \le & \sum_{i=1}^{i^\ast} P_{\tilde{X}^n} ({\bf x}_i) f \left( \Pr \left\{ X^n \in B_n\right\} \right) + 2 \nu \nonumber \\
 & = & f \left( \Pr \left\{ X^n \in B_n\right\} \right) +2 \nu \nonumber \\
 & \le & f \left( f^{-1}(D) \right) + 2 \nu \nonumber \\
 & = & D + 2 \nu
\end{IEEEeqnarray}
for sufficiently large $n$. This completes the proof of the theorem.
\end{IEEEproof}

\begin{theorem} \label{theo:converse}
Under conditions C1) and C2), for any mapping $\phi_n$ satisfying
\begin{equation}
D_f(X^n||\phi_n(U_{M_n})) \le D,
\end{equation}
it holds that
\begin{equation}
\frac{1}{n} \log M_n \ge \frac{1}{n}H_0(1 - f^{-1}(D)|X^n).
\end{equation}
\end{theorem}

\begin{IEEEproof}
It suffices to show the fact that the relation
\begin{equation} \label{eq:b1}
\frac{1}{n} \log M_n < \frac{1}{n}H_0(1 - f^{-1}(D)|X^n)
\end{equation}
necessarily yields
\begin{equation}
D_f(X^n||\phi_n(U_{M_n})) > D.
\end{equation}
We here denote $H' := H_0(1 - f^{-1}(D)|X^n)$ for short.
For any fixed mapping $ \phi_n: {\cal U}_{M_n}\to {\cal X}^n$, we set 
$\tilde{X}^n: = \phi_n(U_{M_n})$ and 
\begin{equation}
B_n := \left\{  {\bf x} \in {\cal X}^n | P_{\tilde{X}^n}({\bf x}) > 0\right\}.
\end{equation}
Then, from the property of the mapping it must hold that
\begin{equation} \label{eq:MB}
M_n \ge |B_n|.
\end{equation}

From the condition C2) the $f$-divergence between $P_{X^n}$ and $P_{\tilde{X}^n}$ is lower bounded by
\begin{IEEEeqnarray}{rCl}
D_f(X^n||\phi_n{(U_{M_n})}) & = & \sum_{{\bf x} \in B_n} P_{\tilde{X}^n}({\bf x}) f\left( \frac{P_{{X}^n}({\bf x})}{P_{\tilde{X}^n}({\bf x})}  \right) \nonumber \\
& \ge &  f\left(  \Pr\{ X^n \in B_n  \}  \right)  \nonumber \\
& \ge & f\left(  \max_{ \substack{B_n \subset {\cal{X}}^n \\ |B_n| \le M_n  }}  \Pr\{ X^n \in B_n  \}  \right)  \nonumber \\
& \ge & f\left(  \max_{ \substack{B_n \subset {\cal{X}}^n \\ \log |B_n| < H'  }}  \Pr\{ X^n \in B_n  \}  \right) \nonumber \\
& > & f\left( 1 - ( 1 - f^{-1}(D)) \right) \nonumber \\
& = & D,
\end{IEEEeqnarray}
where the first inequality is due to (\ref{eq:log-sum}), the second inequality is due to condition C1) and (\ref{eq:MB}) and the third inequality is from (\ref{eq:b1}). The last inequality is from the definition of the alternative expressions given in Theorem \ref{theo:Uye}. This completes the proof.
\end{IEEEproof}
Theorems \ref{theo:direct} and \ref{theo:converse} show that the smooth max entropy and the inverse function of $f$ have important roles in the resolvability problem with respect to $f$-divergences.
From these theorems, we obtain the following theorem which addresses the \textit{general formula} of the optimum resolvability rate.
It should be noted that because of the assumption $0 \le D < f(0)$ and C1),  we have $0 < f^{-1}(D) \le 1$.
\begin{theorem} \label{theo:3-1}
Under conditions C1)--C3), it holds that
\begin{align}
S_r^{(f)}(D|{\bf X}) & = \lim_{\nu \downarrow 0}\limsup_{n \to \infty} \frac{1}{n}H_{0}(1 - f^{-1}(D+\nu)|X^n) \nonumber \\
& = \lim_{\nu \downarrow 0}\limsup_{n \to \infty} \frac{1}{n} H_{0}(1 - f^{-1}(D) + \nu|X^n).
\end{align}
\end{theorem}
\begin{IEEEproof}
We here show the first equality, because the second equality can be derived from the first inequality together with the continuity of the function  $f^{-1}$.

(Direct Part:)
Fix $\nu>0$ arbitrarily.
From Theorem \ref{theo:direct}, for any $\gamma >0$, there exists a mapping $\phi_n$ such that
\begin{IEEEeqnarray}{rCl} \label{eq:t1}
\frac{1}{n} \log M_n
& \le & \frac{1}{n}H_0(1 - f^{-1}(D+\nu)|X^n) +  \gamma, 
\end{IEEEeqnarray}
and
\begin{equation}
D_f(X^n||\phi_n(U_{M_n})) \le D + \nu +  \gamma.
\end{equation}
We here use the diagonal line argument \cite{Han}. Fix a sequence $\{\gamma_i\}_{i=1}^\infty$ such that $\gamma_1 > \gamma_2 > \dots > 0$ and we repeat the above argument as $i \to \infty$.
Then, we can show that there exists a mapping $\phi_n$ satisfying
\begin{equation}
\limsup_{n \to \infty} D_f(X^n||\phi_n(U_{M_n})) \le D + \nu,
\end{equation}
and
\begin{IEEEeqnarray}{rCl} \label{eq:t2}
\limsup_{n \to \infty}\frac{1}{n} \log M_n  \le  \limsup_{n \to \infty}\frac{1}{n}H_0(1 - f^{-1}(D+\nu)|X^n).
\end{IEEEeqnarray}
Here, also from the diagonal line argument with respect to $\nu$, we obtain
\begin{IEEEeqnarray}{rCl} 
\limsup_{n \to \infty}\frac{1}{n} \log M_n   \le \lim_{\nu \downarrow 0} \limsup_{n \to \infty}\frac{1}{n}H_0(1 - f^{-1}(D+\nu)|X^n).
\end{IEEEeqnarray}
This completes the proof of the direct part.

(Converse Part:)
We fixed $\nu >0$ arbitrarily.
From Theorem \ref{theo:converse},  for any mapping $\phi_n$ satisfying
\begin{equation}
D_f(X^n||\phi_n(U_{M_n})) \le D + \nu,
\end{equation}
it holds that
\begin{equation}
\frac{1}{n} \log M_n \ge \frac{1}{n}H_0(1 - f^{-1}(D+\nu)|X^n).
\end{equation}
Consequently, we have
\begin{equation}
\limsup_{n \to \infty}D_f(X^n||\phi_n(U_{M_n})) \le D + \nu
\end{equation}
and
\begin{equation}
\limsup_{n \to \infty}\frac{1}{n} \log M_n \ge \limsup_{n \to \infty} \frac{1}{n}H_0(1 - f^{-1}(D+\nu)|X^n).
\end{equation}
We also use the diagonal line argument \cite{Han}. 
We repeat the above argument as $i \to \infty$ for a sequence $\{\nu_i\}_{i=1}^\infty$ such that $\nu_1 > \nu_2 > \dots > 0$.
Then, for any mapping $\phi_n$ satisfying
\begin{equation}
\limsup_{n \to \infty} D_f(X^n||\phi_n(U_{M_n})) \le D,
\end{equation}
it holds that
\begin{IEEEeqnarray}{rCl}
\limsup_{n \to \infty}\frac{1}{n} \log M_n
&  \ge & \lim_{\nu \downarrow 0}\limsup_{n \to \infty} \frac{1}{n}H_0(1 - f^{-1}(D+\nu)|X^n).
\end{IEEEeqnarray}
This completes the proof of the converse part.
\end{IEEEproof}
\section{Intrinsic randomness problem}
\begin{figure}[h]
\begin{center}
\includegraphics[width=4.5in]{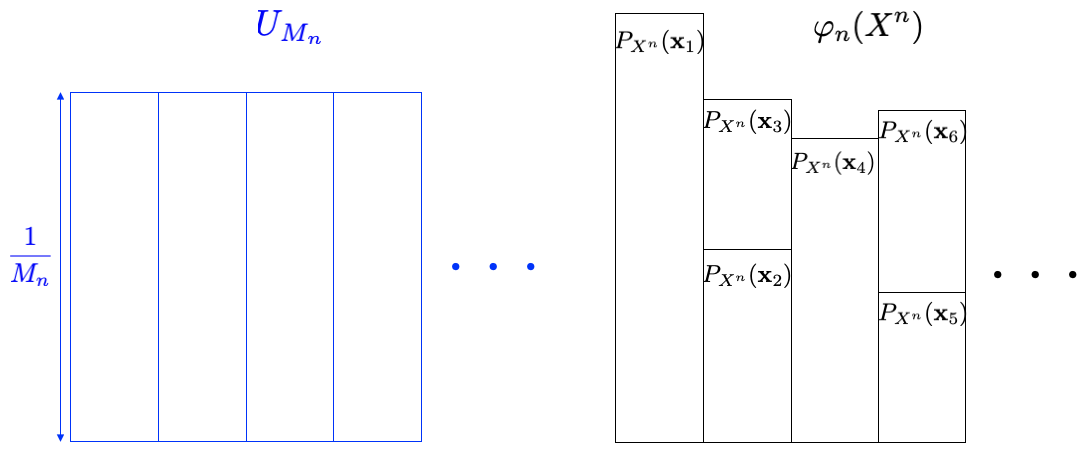}
\centering \caption{Intrinsic randomness problem}
\label{fig:IR}
\end{center}
\end{figure}

In the previous section, we reveal the \textit{general formula} for the optimum resolvability rate.
In this section, we consider how to approximate the uniform random number $U_{M_n}$ by using the given discrete source ${\bf X} = \{ X^n \}_{n=1}^\infty$ and the mapping $\varphi_n$.
Fig. \ref{fig:IR} is an illustrative figure of the problem.
The size of the random number $M_n$ is requested to be as large as possible.
In the intrinsic randomness problem one of our main concern is to derive the largest value of $b$ in the form of $M_n=e^{nb}$ under some approximation measure \cite{VV}.
This problem setting is formulated as follows.
\begin{definition} 
$R$ is said to be $\Delta$-achievable with the given $f$-divergence if there exists a sequence of mapping $\varphi_n : {\cal X}^n \to {\cal U}_{M_n}$ such that
\begin{IEEEeqnarray}{rCl} \label{eq:4-2-1}
\limsup_{n \rightarrow \infty} D_f(\varphi_n(X^n)||U_{M_n}) & \leq & \Delta, \\
\liminf_{n \rightarrow \infty} \frac{1}{n} \log M_n & \geq & R.
\end{IEEEeqnarray}
\end{definition}

In this case, given $\Delta$, if the rate constraint $R$ is sufficiently small, it can be shown that there exists a sequence of mappings that satisfies the constraints. On the other hand, if $R$ is too large, no sequence of mappings that achieves the desired constraints can be found. Consequently, in this setting, the supremum of  $R$ is of particular interest.

\begin{definition}[First-order optimum intrinsic randomness rate] 
\begin{equation} 
S^{(f)}_{\iota}(\Delta|{\bf X}) :=  \sup \left\{ R \left|R \mbox{ is $\Delta$-achievable with the given $f$-divergence} \right. \right\}.
\end{equation}
\end{definition}
\begin{remark}
It should be emphasized that we use the $f$-divergence of the form $D_f(\varphi_n(X^n)||U_{M_n})$ instead of $D_f(U_{M_n}||\varphi_n(X^n))$ (cf. Remark \ref{remark:employ1}).
\end{remark}
We also assume that $\Delta \in [0, f(0))$ in this section (cf. Remark \ref{remark:trivial}).
In order to analyze the general formula of the optimum intrinsic randomness rate $S^{(f)}_{\iota}(\Delta|{\bf X})$, we first give two theorems.
\begin{theorem} \label{theo:4-1}
Under conditions C1) and C2), for any $\gamma >0$ and $M_n$ satisfying 
\begin{equation} \label{eq:assump2}
\frac{1}{n} \log M_n \le \frac{1}{n}H_{\infty}(1 - f^{-1}(\Delta)|X^n) - \gamma,
\end{equation}
there exists a mapping $\varphi_n$ such that
\begin{equation}
D_f(\varphi_n(X^n)||U_{M_n}) \le \Delta + \gamma
\end{equation}
for sufficiently large $n$.
\end{theorem}

\begin{IEEEproof}
We set $\beta_0$ as
\begin{equation}
-\log \beta_0 = H_{\infty}(1 - f^{-1}(\Delta)| {X^n})
\end{equation}
for short.

From Theorem \ref{theo:Uye2}, we notice that 
\begin{equation} \label{eq:ma1}
1 - f^{-1}(\Delta) \ge \sum_{{\bf x} \in {\cal X}^n} \left( P_{X^n}({\bf x}) - \beta_0 \right)^+ =: 1- A_n(\Delta),
\end{equation}
where if $\beta_0 > 1 /|{\cal X}^n|$ holds, then $f^{-1}(\Delta) = A_n(\Delta)$.
We shall show that for any $M_n$ satisfying
\begin{equation} \label{eq:assump2-2-2}
\frac{1}{n} \log M_n \le - \frac{1}{n}\log {\beta_0} - \frac{1}{n}\log \frac{1}{A_n(\Delta)} - \frac{\gamma}{2},
\end{equation}
there exists a mapping $\varphi_n$ such that 
\begin{equation}
D_f(\varphi_n(X^n)||U_{M_n}) \le \Delta + \gamma
\end{equation}
for sufficiently large $n$.

For every sequence ${\bf x} \in {\cal X}^n$, we define the probability distribution
\begin{equation}  \label{eq:a-0-0-1-2}
P_{\overline{X}^n}({\bf x}) := \left\{ \begin{array}{cc}
\frac{\beta_0}{A_n(\Delta)} & P_{{X}^n}({\bf x}) \ge \beta_0, \\
\frac{P_{{X}^n}({\bf x})}{A_n(\Delta)} & P_{{X}^n}({\bf x}) < \beta_0.
\end{array} \right.
\end{equation}
Since $0 < A_n(\Delta) <1$, this probability distribution is well-defined.
Then, from the definition of the smooth min entropy it holds that
\begin{equation}
\sum_{{\bf x} \in {\cal X}^n }P_{\overline{X}^n}({\bf x}) = 1.
\end{equation}
Here, from (\ref{eq:assump2-2-2}) and the definition of the smooth min entropy it holds that
\begin{align}   \label{eq:81-2}
M_n & \le \frac{1}{\beta_0} {A_n(\Delta)} e^{- n\gamma/2} \le |{\cal X}^n|
\end{align}
for sufficiently large $n$.

We next define the mapping $\varphi_n:{\cal X}^n \to {\cal U}_{M_n}$ by using  $P_{\overline{X}^n}$.
To do so, we classify the elements of ${\cal X}^n$ into $I_n(i) \ (1 \le i \le M_n)$ as follows.
\begin{enumerate}
\item We choose a set $I_n(1)$ arbitrarily satisfying
\begin{equation}
\sum_{{\bf x} \in I_n(1)}P_{\overline{X}^n}({\bf x}) \le \frac{1}{M_n},
\end{equation}
\begin{equation}
\sum_{{\bf x} \in I_n(1)}P_{\overline{X}^n}({\bf x}) + P_{\overline{X}^n}({\bf x}') > \frac{1}{M_n}
\end{equation}
for any ${\bf x}' \in {\cal X}^n \setminus I_n(1)$.
\item Next, we choose a set $I_n(2) \subset {\cal X}^n \setminus I_n(1)$ satisfying
\begin{equation}
\sum_{{\bf x} \in I_n(2)}P_{\overline{X}^n}({\bf x}) \le \frac{1}{M_n}, \end{equation}
\begin{equation}
\sum_{{\bf x} \in I_n(2)}P_{\overline{X}^n}({\bf x}) + P_{\overline{X}^n}({\bf x}') > \frac{1}{M_n}
\end{equation}
for any ${\bf x}' \in {\cal X}^n \setminus \bigcup_{i=1}^2 I_n(i)$.
\end{enumerate}
Furthermore, we repeat this operation $(M_n-1)$ times so as to choose sets  $I_n(i) \ (1 \le i \le M_n-1)$.
Notice here that since $\frac{1}{M_n} > \frac{\beta_0}{A_n(\Delta)}$ holds, we can repeat this operation $(M_n-1)$ times. 
Thus, from the above procedure, all of $I_n(i) \ (1 \le i \le M_n-1)$ are not empty.
Lastly,  we set $I_n(M_n) = \{ {\bf x} \in {\cal X}^n | {\bf x} \in {\cal X}^n \setminus \cup_{i=1}^{M_n-1} I_n(i))\}$.

From $I_n(i) \ (1 \le i \le M_n)$, we define the mapping $\varphi_n: {\cal X}^n \to {\cal U}_{M_n}$ as follows:
\begin{equation}
\varphi_n({\bf x}) = i, \quad  {\bf x} \in I_n(i).
\end{equation}
Furthermore, we set $\tilde{U}_{M_n} = \varphi_n(X^n)$. Thus, 
\begin{equation}  \label{eq:a-1-83}
P_{\tilde{U}_{M_n}}(i) = \sum_{{\bf x} \in I_n(i)}P_{{X}^n} ({\bf x})
\end{equation}
holds for every $i$ in $1 \le i \le M_n$.

We next evaluate the above mapping $\varphi_n$.
From the construction of the mapping, it holds that 
\begin{equation} \label{eq:a-1-2}
\frac{1}{M_n}  < \sum_{{\bf x} \in I_n(i)}P_{\overline{X}^n} ({\bf x}) + \frac{\beta_0}{A_n(\Delta)} 
\end{equation}
 for all $i \ (1 \le  i\le M_n-1)$ and
\begin{equation} \label{eq:a-1-1-0-2}
\frac{1}{M_n}  \le \sum_{{\bf x} \in I_n(M_n)}P_{\overline{X}^n} ({\bf x}).
\end{equation}
Hence, for all $i \ (1 \le  i\le M_n)$ we have
\begin{equation} \label{eq:a-1-1-1-2}
\frac{1}{M_n} - \frac{\beta_0}{A_n(\Delta)} \le \sum_{{\bf x} \in I_n(i)}P_{\overline{X}^n} ({\bf x}).
\end{equation}
Here, notice that for all ${\bf x} \in {\cal X}^n$ 
\begin{equation}  \label{eq:a-2-0-2}
P_{\overline{X}^n} ({\bf x}) \le \frac{P_{{X}^n} ({\bf x})}{A_n(\Delta)}
\end{equation}
holds from (\ref{eq:a-0-0-1-2}).
Thus, we have
\begin{IEEEeqnarray}{rCl} \label{eq:a-2-1-2}
\frac{P_{\tilde{U}_{M_n}}( i)}{ A_n(\Delta)}  & = &
\frac{\sum_{{\bf x} \in I_n(i)} P_{{X}^n} ({\bf x})}{A_n(\Delta)} \nonumber \\
& \ge & \sum_{{\bf x} \in I_n(i)} P_{\overline{X}^n} ({\bf x}) \nonumber \\
& > & \frac{1}{M_n}- \frac{\beta_0}{A_n(\Delta)} \nonumber \\
& = & \frac{1}{M_n} \left( 1 - \frac{M_n\beta_0}{ A_n(\Delta)}  \right) \nonumber \\
& \ge & \frac{1}{M_n}  \left( 1 - e^{-n\gamma/2} \right)
\end{IEEEeqnarray}
for all $i \ (1 \le i \le M_n)$ where the first equality is due to (\ref{eq:a-1-83}), the first inequality is due to  (\ref{eq:a-2-0-2}), the second inequality is due to (\ref{eq:a-1-1-1-2}), and the last inequality is due to (\ref{eq:81-2}).
Hence, we obtain
\begin{IEEEeqnarray}{rCl} \label{eq:a-3-1-2}
D_f(\varphi_n(X^n)||U_{M_n}  ) & = & \sum_{1 \le i \le M_n} \frac{1}{M_n} f\left( \frac{P_{\tilde{U}_{M_n}}(i)}{ \frac{1}{M_n}}  \right) \nonumber \\
& \le & \sum_{1 \le i \le M_n} \frac{1}{M_n} f\left(A_n(\Delta) (1 - e^{-n\gamma/2}) \right)  \nonumber \\
& \le & f(f^{-1}(\Delta)) + \delta_n  \nonumber \\
& = & \Delta + \delta_n,
\end{IEEEeqnarray}
where we can choose some $\delta_n >0$ such that $\delta_n \to 0 \ (n \to \infty)$, the first inequality is due to  (\ref{eq:a-2-1-2}), and the second inequality is due to the continuity of the function $f$, (\ref{eq:ma1}) and C1). This completes the proof of the theorem.
\end{IEEEproof}

\begin{theorem} \label{theo:4-2}
Under conditions C1) and C2), for any $\varepsilon >0$ 
if the mapping $\varphi_n$ satisfies 
\begin{equation} \label{eq:18}
D_f(\varphi_n(X^n)||U_{M_n}) \le \Delta - \varepsilon,
\end{equation}
then it holds that
\begin{equation}
\frac{1}{n} \log M_n \le \frac{1}{n}H_{\infty}(1 - f^{-1}(\Delta)|X^n)
\end{equation}
for sufficiently large $n$.
\end{theorem}
\begin{IEEEproof}
Setting
\begin{equation}
H' := H_{\infty}(1 - f^{-1}(\Delta)|X^n),
\end{equation}
we only consider the case where $H' < |{\cal X}^n|$ holds, because $H' = |{\cal X}^n|$ means the trivial result.
Let $\varepsilon >0$ be fixed arbitrarily.
We show that if 
\begin{equation}  \label{eq:cond12}
\frac{1}{n} \log M_n > \frac{1}{n}H' 
\end{equation}
holds for infinitely many $n=n_1,n_2,\dots,$ then for any $\varphi_n$ it holds that
\begin{equation}
D_f(\varphi_n(X^n)||U_{M_n}) > \Delta- \varepsilon.
\end{equation}
From (\ref{eq:cond12}), there exists a positive constant $\gamma$ satisfying
\begin{equation} \label{eq:cond12-2}
\frac{1}{n} \log M_n - 2 \gamma > \frac{1}{n}H'.
\end{equation}
Here, for $\gamma >0$ satisfying the above inequality we set $T_n$ as 
\begin{align}
T_n & := \left\{ {\bf x} \in {\cal X}^n \left| \frac{1}{n} \log \frac{1}{P_{X^n}({\bf x})} \le \frac{1}{n}H' + \gamma \right. \right\} \\
& =  \left\{ {\bf x} \in {\cal X}^n \left| {P_{X^n}({\bf x})} \ge e^{- (H' + n\gamma)} \right. \right\}.
\end{align}
Then, from the relation
\begin{align}
1 \ge \sum_{{\bf x} \in  T_n}  P_{X^n}({\bf x}) \ge |T_n| e^{-(H' + n\gamma) } 
\end{align}
we have
\begin{equation}  \label{eq:P2}
|T_n| \le e^{H' + n\gamma}.
\end{equation}

Next, we fix $M_n$ and a mapping $\varphi_n$ satisfying (\ref{eq:cond12}) and set $\tilde{U}_{M_n}$ as $\tilde{U}_{M_n} = \varphi_n(X^n)$.
Using $\varphi_n$ and $T_n$ we set $I_n$ as
\begin{equation}
I_n := \{i | \exists {\bf x} \in T_n, \varphi_n({\bf x}) = i \}.
\end{equation}
Thus, the set $I_n$ is the set of index $i$ constructing from at least one ${\bf x} \in T_n$ and the set $(I_n)^c$ is the set of $i$ constructing only from ${\bf x} \in (T_n)^c$.

Then, from the definition of the mapping and (\ref{eq:P2}), it holds that
\begin{equation}
|I_n|  \le |T_n| \le  e^{H' + n\gamma}.
\end{equation}
On the other hand, from (\ref{eq:cond12-2}) we have
\begin{align}
{M_n} > e^{H' + 2n \gamma}.
\end{align}
This means that
\begin{equation}   \label{eq:31}
\frac{|I_n|}{M_n} \le e^{-n \gamma}
\end{equation}
holds. 
Hence, from the condition C2) we have
\begin{equation} \label{eq:32}
\frac{|I_n|}{M_n} f\left( \frac{M_n}{|I_n|} \right) \to 0 \quad (n \to \infty).
\end{equation}

From the above argument, the $f$-divergence between $\varphi_n(X^n)$ and $U_{M_n}$ is evaluated as
\begin{IEEEeqnarray}{rCl} \label{eq:68}
D_f(\varphi_n(X^n)||U_{M_n}  ) & = & \sum_{1 \le i \le M_n} \frac{1}{M_n} f\left( \frac{P_{\tilde{U}_{M_n}}(i)}{ \frac{1}{M_n}}  \right) \nonumber \\
& = & \sum_{1 \le i \le M_n, i \in I_n} \frac{1}{M_n} f\left( \frac{P_{\tilde{U}_{M_n}}(i)}{ \frac{1}{M_n}}  \right) \nonumber \\
& & +  \sum_{1 \le i \le M_n, i \in (I_n)^c} \frac{1}{M_n} f\left( \frac{P_{\tilde{U}_{M_n}}(i)}{ \frac{1}{M_n}}  \right) \nonumber \\
& \ge & \frac{|I_n|}{M_n} f\left( \frac{\sum_{1 \le i \le M_n, i \in I_n }P_{\tilde{U}_{M_n}}(i)}{ \frac{|I_n|}{M_n}}  \right)  \nonumber \\
& & +   \frac{|(I_n)^c|}{M_n} f\left( \frac{\sum_{1 \le i \le M_n, i \in (I_n)^c} P_{\tilde{U}_{M_n}}(i)}{ \frac{|(I_n)^c|}{M_n}}  \right) \nonumber \\
& \ge & \frac{|I_n|}{M_n} f\left( \frac{1}{ \frac{|I_n|}{M_n}}  \right) +   \frac{|(I_n)^c|}{M_n} f\left( \frac{\sum_{1 \le i \le M_n, i \in (I_n)^c} P_{\tilde{U}_{M_n}}(i)}{ \frac{|(I_n)^c|}{M_n}}  \right) \nonumber \\
& \ge & \frac{|I_n|}{M_n} f\left( { \frac{M_n}{|I_n|}}  \right) +   \frac{|(I_n)^c|}{M_n} f\left( \frac{\Pr\{X^n \in (T_n)^c\}}{ \frac{|(I_n)^c|}{M_n}}  \right),
\end{IEEEeqnarray}
where the first inequality is due to (\ref{eq:log-sum}), and the last inequality is due to the relation 
\begin{equation}
\sum_{1 \le i \le M_n, i \in (I_n)^c} P_{\tilde{U}_{M_n}}(i) \le \Pr\{X^n \in (T_n)^c \}
\end{equation} and C1).

We next focus on the evaluation of the second term on the RHS of (\ref{eq:68}).
From the definition of the smooth min entropy $H'$ and Theorem \ref{theo:Uye}, for any $\gamma >0$ it necessarily holds that
\begin{IEEEeqnarray}{rCl}
\sum_{{\bf x} \in {\cal X}^n}\left( P_{X^n}({\bf x}) - e^{-(H' + n\gamma)} \right)^+ \ge  1 - f^{-1}(\Delta).
\end{IEEEeqnarray}
Thus, from the definition of $T_n$ it holds that
\begin{IEEEeqnarray}{rCl}
\sum_{{\bf x} \in T_n}\left( P_{X^n}({\bf x}) - e^{-(H' + n\gamma)} \right)
& = &\sum_{{\bf x} \in {\cal X}^n}\left( P_{X^n}({\bf x}) - e^{-(H' + n\gamma)} \right)^+ \nonumber \\
 & \ge & 1 - f^{-1}(\Delta).
\end{IEEEeqnarray}
Thus, we obtain
\begin{IEEEeqnarray}{rCl}
\Pr\left\{  X^n \in T_n \right\} & \ge & 1 - f^{-1}(\Delta),
\end{IEEEeqnarray}
from which it holds that
\begin{IEEEeqnarray}{rCl}
\Pr\left\{  X^n \in (T_n)^c \right\} 
& < 1 - (1- f^{-1}(\Delta))  = f^{-1}(\Delta).
\end{IEEEeqnarray}
Plugging the above inequality with (\ref{eq:68}), we obtain
\begin{IEEEeqnarray}{rCl}
D_f(\varphi_n(X^n)||U_{M_n}  )& > & \frac{|I_n|}{M_n} f\left(  \frac{M_n}{|I_n|}\right) +   \frac{|(I_n)^c|}{M_n} f\left( \frac{f^{-1}(\Delta)}{ \frac{|(I_n)^c|}{M_n}}  \right).
\end{IEEEeqnarray}

Noticing that 
\begin{equation}
\frac{|(I_n)^c|}{M_n} > 1 - e^{-n\gamma},
\end{equation}
from (\ref{eq:31}) for some $\delta_n \to 0$ we have
\begin{IEEEeqnarray}{rCl} \label{eq:637}
D_f(\varphi_n(X^n)||U_{M_n}  )
& > & \left( 1 - e^{- n \gamma} \right) f\left( \frac{f^{-1}(\Delta)}{ 1 - e^{- n \gamma} }  \right) - \delta_n \nonumber \\
& = & f\left( \frac{f^{-1}(\Delta)}{ 1 - e^{- n \gamma} }  \right)  - e^{-n\gamma}f\left( \frac{f^{-1}(\Delta)}{ 1 - e^{- n \gamma} }  \right) - \delta_n \nonumber \\
& = &  f\left( {f^{-1}(\Delta)} (1 + \gamma'_n)  \right) - 2\delta_n
\end{IEEEeqnarray}
for sufficiently large $n$, where we use the property (\ref{eq:32}) and the notation $\gamma'_n = \frac{e^{-n\gamma}}{1 - e^{-n\gamma}}$. 
Since $\gamma'_n \to 0\ (n \to \infty)$ holds, from the continuity of the function $f$
it holds that
\begin{IEEEeqnarray}{rCl} \label{eq:638}
D_f(\varphi_n(X^n)||U_{M_n}  )& > & \Delta - 3\delta_n \nonumber \\
& \ge & \Delta - \varepsilon
\end{IEEEeqnarray}
for $n=n_j, n_{j+1}, \dots, $ with some $j\ge 1$.
Therefore, we obtain the theorem.
\end{IEEEproof}

Theorems \ref{theo:4-1} and \ref{theo:4-2} show that the smooth min entropy and the inverse function of $f$ have important roles in the intrinsic randomness problem with respect to $f$-divergences, while the smooth max entropy is important in the resolvability problem.
By using the above two theorems, we obtain the following theorem.
It should be noted that because of the assumption $0 \le \Delta < f(0)$ and C1),  we have $0 < f^{-1}(\Delta) \le 1$.
\begin{theorem} \label{theo:4-3}
Under conditions C1) and C2), it holds that
\begin{IEEEeqnarray}{rCl}
S^{(f)}_\iota (\Delta|{\bf X}) & = & \lim_{\nu \downarrow 0}\liminf_{n \to \infty} \frac{1}{n}H_{\infty}(1 - f^{-1}(\Delta+\nu)|X^n) \nonumber \\
& = &\lim_{\nu \downarrow 0}\liminf_{n \to \infty} \frac{1}{n} H_{\infty}(1 - f^{-1}(\Delta) + \nu|X^n). \ \ 
\end{IEEEeqnarray}
\end{theorem}
\begin{IEEEproof}

(Direct Part:)
Fix $\nu >0$ arbitrarily. From Theorem \ref{theo:4-1}, for any $\gamma > 0$ and $M_n$ such that 
\begin{align} 
 \frac{1}{n}H_{\infty}(1 - f^{-1}(\Delta + \nu)|X^n) - 2\gamma & \le \frac{1}{n} \log M_n  \le \frac{1}{n}H_{\infty}(1 - f^{-1}(\Delta)|X^n) - \gamma
\end{align}
holds, there exists a mapping $\varphi_n$ satisfying
\begin{equation}
D_f(\varphi_n(X^n)||U_{M_n}) \le \Delta +\gamma
\end{equation}
for sufficiently large $n$.

Since 
$\gamma >0$ is arbitrarily, we obtain
\begin{equation} 
\liminf_{n \to \infty} \frac{1}{n} \log M_n  \ge \liminf_{n \to \infty} \frac{1}{n}H_{\infty}(1 - f^{-1}(\Delta + \nu)|X^n).
\end{equation}

Here, fix a sequence $\{\nu_i\}_{i=1}^\infty$ such that $\nu_1 > \nu_2 > \dots > 0$ and we repeat the above argument as $i \to \infty$.
Then, we can show that there exists a mapping $\varphi_n$ satisfying
\begin{equation} 
\liminf_{n \to \infty}\frac{1}{n} \log M_n  \ge \lim_{\nu \downarrow 0} \liminf_{n \to \infty}\frac{1}{n}H_\infty(1 - f^{-1}(D+\nu)|X^n),
\end{equation}
and
\begin{equation}
\limsup_{n \to \infty }D_f(\varphi_n(X^n)||U_{M_n}) \le \Delta.
\end{equation}
This completes the proof of the direct part of the theorem.

(Converse Part:)
Fix $\nu >0$ arbitrarily. From Theorem \ref{theo:4-2}, 
for any mapping $\varphi_n$ satisfying
\begin{equation}
D_f(\varphi_n(X^n)||U_{M_n}) \le \Delta + \nu,
\end{equation}
it holds that
\begin{equation}
\frac{1}{n} \log M_n \le \frac{1}{n}H_\infty(1 - f^{-1}(\Delta+2\nu)|X^n).
\end{equation}
Thus, for any $\nu >0$, we obtain
\begin{equation}
\limsup_{n \to \infty}D_f(\varphi_n(X^n)||U_{M_n}) \le \Delta + \nu,
\end{equation}
and
\begin{equation}
\liminf_{n \to \infty}\frac{1}{n} \log M_n \le \liminf_{n \to \infty} \frac{1}{n}H_\infty(1 - f^{-1}(\Delta+2\nu)|X^n).
\end{equation}
Noting that $\nu>0$ is arbitrarily, fix a sequence $\{\nu_i\}_{i=1}^\infty$ such that $\nu_1 > \nu_2 > \dots > 0$ and we repeat the above argument as $i \to \infty$.
Then, we obtain
\begin{equation}
\limsup_{n \to \infty} D_f(\varphi_n(X^n)||U_{M_n}) \le \Delta,
\end{equation}
and
\begin{align}
\liminf_{n \to \infty}\frac{1}{n} \log M_n \le \lim_{\nu \downarrow 0}\liminf_{n \to \infty} \frac{1}{n}H_\infty(1 - f^{-1}(\Delta+\nu)|X^n).
\end{align}
This completes the proof of the converse part of the theorem.
\end{IEEEproof}
%
%
%
%
%
%
%
%
\section{Relaxation of Conditions C1) and C2)} \label{sect:relaxed_condition}
So far, we have derived the general formulas for the optimum resolvability rate under conditions C1)--C3) and for the optimum intrinsic randomness rate under conditions C1) and C2). 
In this section, we relax conditions C1) and C2) to extend the class of $f$-divergence for which we can characterize these optimum rates. Hereafter, we do not consider a linear function $f(t) = a (t-1)$ with some $a$ because it always gives a trivial case where $D_f(Z || \overline{Z}) = 0$. 
 
We consider the following condition, which is a relaxation of C2):
\begin{itemize}
\item[C2')] The function $f$ satisfies
\begin{align}
\lim_{u \rightarrow \infty} \frac{f(u)}{u} < + \infty. \label{eq:101}
\end{align}
\end{itemize}
For function $f$ satisfying condition C2'), we denote the LHS of \eqref{eq:101} by 
\begin{align}
c_f = \lim_{u \rightarrow \infty} \frac{f(u)}{u}.
\end{align}
We give some examples of the function $f(t)$ which satisfies C2') but not C2).
\begin{itemize}
\item $f(t) = |t-1|$: The $f$-divergence is variational distance, and $c_f=1$. 
\item $f(t) = (1- \sqrt{t})^2$: The $f$-divergence is squared Hellinger distance, and $c_f = 1$.
\item $f(t) = \frac{t^\alpha - \alpha t -(1-\alpha)}{\alpha(\alpha-1)}$ ($0 < \alpha <1$): The $f$-divergence is $\alpha$-divergence, and $c_f = \frac{1}{1-\alpha}$.
\end{itemize}

For function $f(t)$ satisfying condition C2'), we consider its modified function 
\begin{align}
f_{0} (t) := f(t) + c_f(1-t), \label{eq:offset_function}
\end{align}
 which is offset by $c_f(1-t)$. This function is called the offset function of $f$.
It should be noted that under condition C2), which is a special case of C2'), it holds that $c_f = 0$ and thus $f_0(t) = f(t)$ for all $t \ge 0$. 
We have the following lemma:
\begin{lemma} \label{lem:offset_f}
Assume that the function $f(t)$ satisfies condition C2'). Then, 
\begin{enumerate}
\item[(i)] the offset function $f_0$ satisfies conditions C1) and (C2),
\item[(ii)] for any pair of probability distributions $P_Z$ and $P_{\overline{Z}}$ with the same alphabet $\mathcal{Z}$, it holds  that 
\begin{align}
D_f(Z || \overline{Z}) = D_{f_0} (Z || \overline{Z}).
\end{align}
\end{enumerate}
\end{lemma}
\begin{IEEEproof} 
It is easily verified that $f_0$ is a convex function with $f_0(1) = 0$, and claim (ii) is well-known. So, here we show claim (i). By definition, it holds that
\begin{align}
\lim_{u \rightarrow \infty} \frac{f_0(u)}{u} &= \lim_{u \rightarrow \infty}\left( \frac{f(u)}{u} + \frac{c_f(1-u)}{u} \right) \nonumber \\
&= \lim_{u \rightarrow \infty} \frac{f(u)}{u} - c_f = 0, \label{eq:C2_holding}
\end{align}
which indicates that $f_0$ satisfies condition C2).

To show  C1) being hold, we  use the left-derivative of $f_0$ at $t > 0$, denoted as
\begin{align}
f_0'(t-) = \lim_{h \uparrow 0} \frac{f_0(t+h) - f_0(t)}{h}
\end{align}
(cf. \cite{Rockafellar}).
Contrary to ordinary derivatives, the left-derivative at $t > 0$ always exists for function $f_0$, which is continuous.
To show that $f_0$ satisfies condition C1), it suffices to show that$f_0'(t-) \le 0$ for all $t>0$. 
Using the left-derivative $f'_0(t-)$, a tangent line at $t>0$ can be expressed as $f_0'(t-) \cdot t + b$ with some $b$, where $f_0'(t-)$ and $b$ correspond to the slope and intercept of this tangent line, respectively. 
We call this tangent line the  left-tangent line at $t$.
Fixing $t^* > 0$ arbitrarily, let $a^* := f_0'(t^*-)$ and $b^*$ be the intercept of the left-tangent line at $t^*$. The convexity of $f_0$ implies that 
\begin{align}
f_0(t) \ge a^* t + b^*~~(\forall t \ge 0).
\end{align} 
Then, it follows from \eqref{eq:C2_holding} that
\begin{align}
0 = \lim_{u \rightarrow \infty} \frac{f_0(u)}{u} \ge \lim_{u \rightarrow \infty} \frac{a^* u + b^*}{u} = a^* = f_0'(t^*-).
\end{align}
Since $t^* > 0$ is arbitrary, this inequality implies that $f_0(t)$ is decreasing for $t>0$ with $f_0(0)>0$, completing the proof of the lemma.
\end{IEEEproof}

Lemma \ref{lem:offset_f} indicates that if the original function $f$ satisfies condition C2'), then its offset function $f_0$ satisfies conditions C1) and C2) without changing the value of $f$-divergence. 
Because condition C2) is a special instance of condition C2') with $c_f = 0$, claim (i) of Lemma \ref{lem:offset_f} implies that condition C1) is superfluous for functions satisfying C2) (cf.\ Remark \ref{remark:relaxed_condition}). 
The following proposition is immediately obtained by {claim} (ii) of Lemma \ref{lem:offset_f}: 
\begin{proposition} \label{prop:optimum_rates}
Assume that the function $f(t)$ satisfies condition C2'). Then, we have
\begin{align}
S_r^{(f)}(D|{\bf X}) &= S_r^{(f_0)}(D|{\bf X}), \\
S^{(f)}_\iota (\Delta|{\bf X}) &= S^{(f_0)}_\iota (\Delta|{\bf X}).
\end{align}
\end{proposition}

It is easily verified if $f$ satisfies condition C3) as well as C2'), then so does $f_0$. 
From this fact, Lemma \ref{lem:offset_f}, and Proposition \ref{prop:optimum_rates}, we have the following generalization of Theorems \ref{theo:3-1}.
\begin{theorem} \label{theo:general3-1}
Under conditions C2') and C3), it holds that
\begin{align}
S_r^{(f)}(D|{\bf X}) &= \lim_{\nu \downarrow 0}\limsup_{n \to \infty} \frac{1}{n}H_{0}(1 - f_0^{-1}(D+\nu)|X^n) \nonumber \\
& = \lim_{\nu \downarrow 0}\limsup_{n \to \infty} \frac{1}{n} H_{0}(1 - f_0^{-1}(D) + \nu|X^n).
\end{align}
\end{theorem}

For the optimum intrinsic randomness rate, we also have the generalized result of Theorem \ref{theo:4-3}. 
\begin{theorem} \label{theo:general4-3}
Under condition C2'), it holds that
\begin{align}
S_\iota^{(f)}(D|{\bf X}) &= \lim_{\nu \downarrow 0}\liminf_{n \to \infty} \frac{1}{n}H_{\infty}(1 - f_0^{-1}(\Delta+\nu)|X^n) \nonumber \\
& = \lim_{\nu \downarrow 0}\liminf_{n \to \infty} \frac{1}{n} H_{\infty}(1 - f_0^{-1}(\Delta) + \nu|X^n).
\end{align}
\end{theorem}
%
%
%
%
%
%
%
\section{Particularization to several distance measures}
In previous sections, we have derived the general formula of the \textit{first-order} optimum resolvability and intrinsic randomness rates with respect to $f$-divergences, where the smooth R\'enyi entropy and the inverse function of $f$ have important roles.
In this section, we  first focus on several specified functions $f$ satisfying {conditions C1)--C3)} and 
compute these rates by using Theorems \ref{theo:3-1} and \ref{theo:4-3}.
In addition, we consider the function $f$ satisfying C2') and C3) and compute the rates by using Theorems \ref{theo:general3-1} and \ref{theo:general4-3}.
It will turn out that it is easy to derive the optimum achievable rates for specified approximation measures.
We use the notation
\begin{IEEEeqnarray}{rCl}
D_f(X^n||\tilde{X}^n) & := & D_f(X^n||\phi_n(U_{M_n})), \quad D_f(\tilde{U}_{M_n}||U_{M_n}) := D_f(\varphi_n(X^n)||U_{M_n})
\end{IEEEeqnarray}
for convenience.
\begin{remark}  \label{remark:5-1}
Since the function $f(t) = t \log t$ (which indicates the KL divergence) does not satisfy C1) and C2), we can not apply Theorems \ref{theo:3-1} and \ref{theo:4-3}  to the case of the KL divergence: 
\begin{IEEEeqnarray}{rCl}  \label{eq:5KL}
D_f(X^n||\tilde{X}^n) = D(X^n||\tilde{X}^n) &= & \sum_{{\bf x} \in {\cal X}^n} P_{{X}^n}({\bf x}) \log \frac{P_{{X}^n}({\bf x})}{P_{\tilde{X}^n}({\bf x})},  \\
D_f(\tilde{U}_{M_n}||U_{M_n}) = D(\tilde{U}_{M_n}||U_{M_n}) &= & \sum_{1 \le i \le M_n} P_{\tilde{U}_{M_n}}(i) \log \frac{P_{\tilde{U}_{M_n}}(i)}{P_{{U}_{M_n}}(i)}.
\end{IEEEeqnarray} 
The resolvability problem with respect to the KL divergence of this direction has not been considered yet.
On the other hand, in the intrinsic randomness problem, Hayashi \cite[Theorem 7]{Hayashi} has studied the problem with respect to the normalized KL divergence: $1/nD(\tilde{U}_{M_n}||U_{M_n})$ as well as $D(U_{M_n}||\tilde{U}_{M_n})$.
\end{remark}

\subsection{Half variational distance}
We first consider the case of $f(t)$ given as $f(t) = (1-t)^+$ which indicates 
\begin{IEEEeqnarray}{rCl}
D_f(X^n||\tilde{X}^n) & = & \frac{1}{2}\sum_{{\bf x} \in {\cal X}^n} \left| P_{\tilde{X}^n}({\bf x}) - P_{X^n}({\bf x}) \right|, \\
D_f(\tilde{U}_{M_n}||U_{M_n}) & = &  \frac{1}{2}\sum_{1 \le i \le M_n} \left| P_{\tilde{U}_{M_n}}(i) - P_{U_{M_n}}(i) \right|.
\end{IEEEeqnarray}
In this special case, we obtain the following corollary:
\begin{corollary}  
\label{coro:vd}
For $f(t) = (1-t)^+$, it holds that
\begin{IEEEeqnarray}{rCl} S_r^{(f)}(D|{\bf X}) &=& \lim_{\nu \downarrow 0}\limsup_{n \to \infty} \frac{1}{n}H_{0}(D+\nu|X^n),  \\
{S}_\iota^{(f)}({\Delta}|{\bf X}) & = &  \lim_{\nu \downarrow 0}\liminf_{n \to \infty} \frac{1}{n}H_{\infty}(\Delta+\nu|X^n).
\end{IEEEeqnarray}
\end{corollary}
\begin{IEEEproof}
In the case of $f(t) = (1-t)^+$, the inverse function becomes $f^{-1}(D)= 1 - D$, because $0 \le D <1$ holds.
Hence, from Theorems  \ref{theo:3-1} and \ref{theo:4-3} we obtain the corollary. 
\end{IEEEproof}

The former result in the above corollary coincides with the result given by Uyematsu \cite[Theorem 6]{Uyematsu_ISIT2010} while the latter one coincides with the result given by Uyematsu and Kunimatsu\cite[Theorem 6]{Uyematsu13}.
%
%
\subsection{Reverse Kullback-Leibler divergence}
Secondly, we consider the case of $f(t) = -\log t$, which indicates
\begin{IEEEeqnarray}{rCl}
D_f(X^n||\tilde{X}^n) & = & D(\phi_n(U_{M_n})||X^n) = \sum_{{\bf x} \in {\cal X}^n} P_{\tilde{X}^n}({\bf x}) \log \frac{P_{\tilde{X}^n}({\bf x})}{P_{{X}^n}({\bf x})}, \\
D_f(\tilde{U}_{M_n}||U_{M_n}) & = & D(U_{M_n} || \varphi_n(X^n)) = \sum_{1 \le i \le M_n} P_{U_{M_n}}(i) \log \frac{P_{U_{M_n}}(i)}{P_{\tilde{U}_{M_n}}(i)} .
\end{IEEEeqnarray}

In this case, we obtain the following corollary:
\begin{corollary} 
For $f(t) = -\log t$, it holds that
\begin{IEEEeqnarray}{rCl} S_r^{(f)}(D|{\bf X}) &=& \lim_{\nu \downarrow 0}\limsup_{n \to \infty} \frac{1}{n}H_{0}(1 - e^{-(D+\nu)}|X^n),  \\
{S}_\iota^{(f)}({\Delta}|{\bf X}) & = &   \lim_{\nu \downarrow 0}\liminf_{n \to \infty} \frac{1}{n}H_{\infty}(1 - e^{-(\Delta+\nu)}|X^n).
\end{IEEEeqnarray}
\end{corollary}
\begin{IEEEproof}
The inverse function is immediately given by $f^{-1}(D)= e^{-D}$.
Hence, from Theorems  \ref{theo:3-1} and \ref{theo:4-3} we obtain the corollary. 
\end{IEEEproof}
%
%
%
%
\subsection{Hellinger distance}
We consider the case of $f(t) = 1 - \sqrt{t}$, which indicates
\begin{IEEEeqnarray}{rCl}
D_f(X^n||\tilde{X}^n) & = & 1- \sum_{{\bf x} \in {\cal X}^n}\sqrt{P_{X^n}({\bf x}) P_{\tilde{X}^n}({\bf x})}, \\
D_f(\tilde{U}_{M_n}||U_{M_n}) & = &  1- \sum_{1 \le i \le M_n} \sqrt{ P_{\tilde{U}_{M_n}}(i) P_{U_{M_n}}(i) }.
\end{IEEEeqnarray}

In this case, we obtain the following corollary:
\begin{corollary} 
For $f(t) =  1-\sqrt{t}$, it holds that
\begin{IEEEeqnarray}{rCl} S_r^{(f)}(D|{\bf X}) &=& \lim_{\nu \downarrow 0}\limsup_{n \to \infty} \frac{1}{n}H_{0}(2D-D^2+\nu|X^n),  \\
{S}_\iota^{(f)}({\Delta}|{\bf X})  &=& \lim_{\nu \downarrow 0}\liminf_{n \to \infty} \frac{1}{n}H_{\infty}(2\Delta-\Delta^2+\nu|X^n).
\end{IEEEeqnarray}
\end{corollary}
\begin{IEEEproof}
The inverse function of $f(t) =  1-\sqrt{t}$ is given by $f^{-1}(D)= (1 - D)^2$.
Hence, from Theorems  \ref{theo:3-1} and \ref{theo:4-3} we obtain the corollary. Notice here that since both of $D$ and $\Delta$ are smaller than one, $2D-D^2$ as well as $2\Delta-\Delta^2$ are positive.
\end{IEEEproof}
\subsection{$E_\gamma$-divergence}
We consider the case of $f(t) = (\gamma - t)^+ + 1- \gamma $, which indicates
\begin{IEEEeqnarray}{rCl}
D_f(X^n||\tilde{X}^n) 
& = & \sum_{{\bf x} \in {{\cal X}^n}: P_{{X^n}}({\bf x}) > \gamma P_{\tilde{X}^n}({\bf x})} \left(P_{{X^n}}({\bf x}) - \gamma P_{\tilde{X}^n}({\bf x}) \right). \\
D_f(\tilde{U}_{M_n}||U_{M_n})
& = &  \sum_{1 \le i \le M_n : P_{\tilde{U}_{M_n}}(i) > \gamma P_{U_{M_n}}(i) } \left(P_{\tilde{U}_{M_n}}(i)  - \gamma P_{{U}_{M_n}}(i) \right). 
\end{IEEEeqnarray}

In this case, we obtain the corollary:
\begin{corollary} 
 \label{coro:egamma}
For $f(t) =  (\gamma - t)^+ + 1- \gamma$, we have
\begin{IEEEeqnarray}{rCl} S_r^{(f)}(D|{\bf X}) &=& \lim_{\nu \downarrow 0}\limsup_{n \to \infty} \frac{1}{n}H_{0}(D+\nu|X^n),  \\
{S}_\iota^{(f)}({\Delta}|{\bf X}) & = &  \lim_{\nu \downarrow 0}\liminf_{n \to \infty} \frac{1}{n}H_{\infty}(\Delta+\nu|X^n).
\end{IEEEeqnarray}
\end{corollary}
\begin{IEEEproof}
Noting that $\gamma \ge 1$, we have $f(t) = 1 -t$. Hence, the corollary holds.
\end{IEEEproof}
\begin{remark}  \label{remark:5-3}
The above corollary shows that both of optimum achievable rates with respect to the $E_\gamma$-divergence does not depend on $\gamma$, which means that these rates coincides with the optimum achievable rates with respect to the half variational distance (cf. Corollary \ref{coro:vd}).
\end{remark}
%
%
\subsection{Variational distance}
We next consider functions $f$ satisfying C2') and C3). 
Firstly, the function $f(t) = |1-t|$ is considered:
\begin{IEEEeqnarray}{rCl}
D_f(X^n||\tilde{X}^n) & = & \sum_{{\bf x} \in {\cal X}^n} \left| P_{\tilde{X}^n}({\bf x}) - P_{X^n}({\bf x}) \right|, \\
D_f(\tilde{U}_{M_n}||U_{M_n}) & = &  \sum_{1 \le i \le M_n} \left| P_{\tilde{U}_{M_n}}(i) - P_{U_{M_n}}(i) \right|.
\end{IEEEeqnarray}
As we have already mentioned in the previous section, $f(t) = |1-t|$ does not satisfy C1). However it satisfies C2') and C3). Hence, from Theorems \ref{theo:general3-1} and \ref{theo:general4-3}, we obtain the corollary:
\begin{corollary} 
For $f(t) =  |1- t|$, we have
\begin{IEEEeqnarray}{rCl} S_r^{(f)}(D|{\bf X}) &=& \lim_{\nu \downarrow 0}\limsup_{n \to \infty} \frac{1}{n}H_{0}\left( \left. \frac{D}{2} + \nu \right| X^n \right),  \\
{S}_\iota^{(f)}({\Delta}|{\bf X}) & = &  \lim_{\nu \downarrow 0}\liminf_{n \to \infty} \frac{1}{n}H_{\infty}\left( \left. \frac{\Delta}{2} + \nu \right| X^n \right).
\end{IEEEeqnarray}
\end{corollary}
\begin{IEEEproof}
Noticing that $c_f = 1$, we have $f_0(t) = |1-t| + (1-t)$, from which we obtain
\begin{equation}
f^{-1}_0(D) = 1 - \frac{D}{2}.
\end{equation}
Therefore, we obtain the corollary from Theorems \ref{theo:general3-1} and \ref{theo:general4-3}.
\end{IEEEproof}
\subsection{Squared Hellinger distance}
We consider the function $f(t) = ( 1- \sqrt{t})^2$, which is also satisfies C2') and C3). It indicates
\begin{IEEEeqnarray}{rCl}
D_f(X^n||\tilde{X}^n) & = & \sum_{{\bf x} \in {\cal X}^n} \left( \sqrt{P_{\tilde{X}^n}({\bf x})} - \sqrt{P_{X^n}({\bf x})} \right)^2, \\
D_f(\tilde{U}_{M_n}||U_{M_n}) & = &  \sum_{1 \le i \le M_n} \left( \sqrt{ P_{\tilde{U}_{M_n}}(i)} - \sqrt{P_{U_{M_n}}(i)} \right)^2.
\end{IEEEeqnarray}
In this case, we also apply Theorems \ref{theo:general3-1} and \ref{theo:general4-3}.
\begin{corollary} 
For $f(t) =  ( 1- \sqrt{t})^2$, we have
\begin{IEEEeqnarray}{rCl} S_r^{(f)}(D|{\bf X}) &=& \lim_{\nu \downarrow 0}\limsup_{n \to \infty} \frac{1}{n}H_{0}\left( \left. D - \frac{D^2}{4} + \nu \right| X^n \right),  \\
{S}_\iota^{(f)}({\Delta}|{\bf X}) & = &  \lim_{\nu \downarrow 0}\liminf_{n \to \infty} \frac{1}{n}H_{\infty}\left( \left. D - \frac{D^2}{4} + \nu \right| X^n \right).
\end{IEEEeqnarray}
\end{corollary}
\begin{IEEEproof}
Noticing that $c_f = 1$, we obtain
\begin{equation}
f^{-1}_0(D) = \left(1 - \frac{D}{2} \right)^2.
\end{equation}
Hence, we obtain the corollary.
\end{IEEEproof}
\subsection{$\alpha$-divergence}
We consider the function $f(t) = \frac{t^\alpha - \alpha t -(1-\alpha)}{\alpha(\alpha-1)}$ ($0 < \alpha <1$), which is also satisfies C2') and C3). The $\alpha$-divergence in our setting is given by
\begin{IEEEeqnarray}{rCl}
D_f(X^n||\tilde{X}^n) & = & \frac{1}{\alpha(1-\alpha)}\left( 1 - \sum_{{\bf x} \in {\cal X}^n} {P_{X^n}({\bf x})^{\alpha}}{P_{\tilde{X}^n}({\bf x})^{1-\alpha}} \right), \\
D_f(\tilde{U}_{M_n}||U_{M_n}) & = &   \frac{1}{\alpha(1-\alpha)} \left( 1 - \sum_{1 \le i \le M_n} P_{\tilde{U}_{M_n}}(i)^{\alpha}{P_{U_{M_n}}(i)^{1-\alpha}} \right).
\end{IEEEeqnarray}
In this case, we obtain the following corollary using Theorems \ref{theo:general3-1} and \ref{theo:general4-3}.
\begin{corollary} 
For $f(t) =  \frac{t^\alpha - \alpha t -(1-\alpha)}{\alpha(\alpha-1)}$, we have
\begin{IEEEeqnarray}{rCl} S_r^{(f)}(D|{\bf X}) &=& \lim_{\nu \downarrow 0}\limsup_{n \to \infty} \frac{1}{n}H_{0}\left( \left. 1 - \left(D \alpha(\alpha-1) +1 \right)^{1/\alpha} + \nu \right| X^n \right),  \\
{S}_\iota^{(f)}({\Delta}|{\bf X}) & = &  \lim_{\nu \downarrow 0}\liminf_{n \to \infty} \frac{1}{n}H_{\infty}\left( \left. 1-\left(D \alpha(\alpha-1) +1 \right)^{1/\alpha} + \nu \right| X^n \right).
\end{IEEEeqnarray}
\end{corollary}
\begin{IEEEproof}
Noticing that $c_f = 1/(1-\alpha)$, we obtain
\begin{IEEEeqnarray}{rCl}
f_0(t) & = &  \frac{t^\alpha -1 }{\alpha(\alpha-1)} \iff  f^{-1}_0(D) = \left(D \alpha(\alpha-1) +1 \right)^{1/\alpha}.
\end{IEEEeqnarray}
Hence, we obtain the corollary.
\end{IEEEproof}
Let us consider the case of $\alpha = 1/2$. In this case, the inverse function can be simply expressed as 
\begin{equation}
 f^{-1}_0(D) = \left( 1 - \frac{D}{4} \right)^{2}.
\end{equation}
Hence, we have 
\begin{IEEEeqnarray}{rCl} S_r^{(f)}(D|{\bf X}) &=& \lim_{\nu \downarrow 0}\limsup_{n \to \infty} \frac{1}{n}H_{0}\left( \left. \frac{D}{2} - \frac{D^2}{16} + \nu \right| X^n \right).
\end{IEEEeqnarray}
It is known that $\alpha$-divergence with $\alpha = 1/2$ is related to the squared Hellinger distance.
In actual, the optimum resolvability rate  $S_r^{(f)}(D|{\bf X})$ with respect to the squared Hellinger distance is identical to $S_r^{(f)}(2D|{\bf X})$ with respect to the $\alpha$-divergence with $\alpha =1/2$.
%
%
%
%
%
%
\section{Second-order optimum achievable rate}
%
%
%
So far, we have considered the \textit{first-order} optimum resolvability rate as well as the \textit{first-order} optimum intrinsic randomness rate.
The rate of the \textit{second-order}, which enables us to make a finer evaluation of achievable rates, has already been investigated in several information theoretic problems  \cite{Hayashi,Hayashi2,Poly2010,Ingber2011,NH2011,KV2012,KV2013,TK2014,NH2014,YHN2016,W2017}.
In this section, according to these results we also consider the \textit{second-order} optimum achievable rates in random number generation problems with respect to $f$-divergences.
\subsection{General Formula}
We first define the second-order achievability in the resolvability problem.
\begin{definition} 
{\emph L} is said to be $(D, R)$-achievable with the given $f$-divergence if there exists a sequence mapping $\phi_n : {\cal U}_{M_n} \to {\cal X}^n$ such that 
\begin{IEEEeqnarray}{rCl}
\limsup_{n \rightarrow \infty} D_f(X^n||\phi_n(U_{M_n})) \leq D, \\
\limsup_{n \rightarrow \infty} \frac{1}{\sqrt{n}}\log \frac{M_n}{e^{nR}} \leq L.
\end{IEEEeqnarray}
\end{definition}
\begin{definition} [Second-order optimum resolvability rate]
\begin{eqnarray*} 
S_r^{(f)}(D,R|{\bf X}) := \inf \left\{ L \left|L \mbox{ is $(D, R)$-achievable with the given $f$-divergence} \right. \right\}.
\end{eqnarray*}
\end{definition}
In order to analyze the above quantity, we use the following condition instead of C3):
\begin{description}
\item[C3')] For any pair of positive real numbers $(a,b)$, it holds that
\begin{equation}
\lim_{n\to \infty} \frac{f\left(e^{-{n}b} \right)}{e^{\sqrt{n}a}} = 0.
\end{equation}
\end{description}
Here, functions $f(t) = -\log t$, $f(t) = 1 - \sqrt{t}$, $f(t) = ( 1- t)^+$, and $f(t) = (\gamma- t)^+ + ( 1 - \gamma)$ satisfy the condition C3').
Then, the following theorem holds:
\begin{theorem}[Second-order optimum resolvability rate] \label{theo:5-1}
Under conditions C2') and C3'), it holds that
\begin{align}
S_r^{(f)}(D,R|{\bf X}) & = \lim_{\nu \downarrow 0}\limsup_{n \to \infty} \frac{H_{0}(1 - f_0^{-1}(D+\nu)|X^n) - nR}{\sqrt{n}}, \label{eq:2nd_order_R_rate1} 
\end{align}
where $f_0$ is the offset function of $f$, defined in \eqref{eq:offset_function}.

In particular, under conditions C2) and C3'), it holds that
\begin{align}
S_r^{(f)}(D,R|{\bf X}) & = \lim_{\nu \downarrow 0}\limsup_{n \to \infty} \frac{H_{0}(1 - f^{-1}(D+\nu)|X^n) - nR}{\sqrt{n}}. \label{eq:2nd_order_R_rate2} 
\end{align}
\end{theorem}

\begin{IEEEproof} 
Noticing that Lemma \ref{lem:offset_f} indicates the offset function $f_0$ satisfies conditions C1)--C3), the proof of \eqref{eq:2nd_order_R_rate1} proceeds in parallel with proofs of Theorems \ref{theo:direct}, \ref{theo:converse}, and \ref{theo:3-1} in which $f$, $\frac{1}{n}$ and $e^{-n\gamma}$ are replaced by $f_0$, $\frac{1}{\sqrt{n}}$ and  $e^{-\sqrt{n}\gamma}$, respectively.
Equation \eqref{eq:2nd_order_R_rate2} is a special case of \eqref{eq:2nd_order_R_rate1} with $f_0 = f$.
\end{IEEEproof}
%
%
%
We next consider the case of the intrinsic randomness problem. 
\begin{definition} 
$L$ is said to be $(\Delta,R)$-achievable with the given $f$-divergence if there exists a sequence of mapping $\varphi_n : {\cal X}^n \to {\cal U}_{M_n}$ such that
\begin{align} \label{eq:6-2-1}
\limsup_{n \rightarrow \infty} D_f(\varphi_n(X^n)||U_{M_n}) & \leq \Delta, \\
\liminf_{n \rightarrow \infty} \frac{1}{\sqrt{n}} \log \frac{M_n}{e^{nR}} & \geq L.
\end{align}
\end{definition}
\begin{definition}[Second-order optimum intrinsic randomness rate] 
\begin{equation}
S^{(f)}_{\iota}(\Delta,R|{\bf X}) :=  \sup \left\{ L \left|L \mbox{ is $(\Delta,R)$-achievable with the given $f$-divergence} \right. \right\}. 
\end{equation}
\end{definition}
Then, we have the theorem:
\begin{theorem}[Second-order optimum intrinsic randomness rate] \label{theo:5-2}
Under condition C2'), it holds that
\begin{align}
S^{(f)}_{\iota}(\Delta,R|{\bf X}) = \lim_{\nu \downarrow 0}\liminf_{n \to \infty} \frac{H_{\infty}(1 - f_0^{-1}(\Delta+\nu)|X^n)  - n R}{\sqrt{n}}, \label{eq:2nd_order_IR_rate1}
\end{align}
where $f_0$ is the offset function of $f$.

In particular, under condition C2), it holds that
\begin{align}
S^{(f)}_{\iota}(\Delta,R|{\bf X}) = \lim_{\nu \downarrow 0}\liminf_{n \to \infty} \frac{H_{\infty}(1 - f^{-1}(\Delta+\nu)|X^n)  - n R}{\sqrt{n}}, \label{eq:2nd_order_IR_rate2}
\end{align}
\end{theorem}
\begin{IEEEproof}
The proof of \eqref{eq:2nd_order_IR_rate1} proceeds in parallel with proofs of Theorems \ref{theo:4-1}, \ref{theo:4-2}, and \ref{theo:4-3} in which $f$ and $\frac{1}{n}$ are replaced by $f_0$ and $\frac{1}{\sqrt{n}}$. 
Equation \eqref{eq:2nd_order_IR_rate2} is a special case of \eqref{eq:2nd_order_IR_rate1} with $f_0 = f$.
\end{IEEEproof}
Theorems \ref{theo:5-1} and \ref{theo:5-2} show that in both of the resolvability and the intrinsic randomness the smooth R\'enyi entropy and the inverse function of $f$ have also essential roles to express \textit{second-order} optimum achievable rates.
%
%
%
%
\subsection{Particularizations to several distance measures}
Analogously to Section IV, we compute $S_r^{(f)}(D,R|{\bf X})$ and $S_\iota^{(f)}(\Delta,R|{\bf X})$ for the specified function $f$ satisfying C1), C2) and C3'), by using Theorems \ref{theo:5-1} and \ref{theo:5-2}.
We obtain the following corollary:

\begin{corollary} 
\label{coro:vd2}
For $f(t) = (1-t)^+$, it holds that
\begin{IEEEeqnarray}{rCl}  \label{eq:rvd2}
 S_r^{(f)}(D,R|{\bf X}) &=& \lim_{\nu \downarrow 0}\limsup_{n \to \infty} \frac{H_{0}(D+\nu|X^n) - nR}{\sqrt{n}},  \\ \label{eq:ivd2}
{S}_\iota^{(f)}({\Delta},R|{\bf X}) & = &  \lim_{\nu \downarrow 0}\liminf_{n \to \infty} \frac{H_{\infty}(\Delta+\nu|X^n)-nR}{\sqrt{n}}.
\end{IEEEeqnarray}
For $f(t) = -\log t$, it holds that
\begin{IEEEeqnarray}{rCl} S_r^{(f)}(D,R|{\bf X}) &=& \lim_{\nu \downarrow 0}\limsup_{n \to \infty} \frac{H_{0}(1 - e^{-(D+\nu)}|X^n)-nR}{\sqrt{n}},  \\
{S}_\iota^{(f)}({\Delta},R|{\bf X}) & = &   \lim_{\nu \downarrow 0}\liminf_{n \to \infty} \frac{H_{\infty}(1 - e^{-(\Delta+\nu)}|X^n)-nR}{\sqrt{n}}.
\end{IEEEeqnarray}
For $f(t) =  1-\sqrt{t}$, it holds that
\begin{IEEEeqnarray}{rCl} S_r^{(f)}(D,R|{\bf X}) &=& \lim_{\nu \downarrow 0}\limsup_{n \to \infty} \frac{H_{0}(2D-D^2+\nu|X^n) - nR}{\sqrt{n}},  \\
{S}_\iota^{(f)}({\Delta},R|{\bf X})  &=& \lim_{\nu \downarrow 0}\liminf_{n \to \infty} \frac{H_{\infty}(2\Delta-\Delta^2+\nu|X^n) - nR}{\sqrt{n}}.
\end{IEEEeqnarray}
 \label{coro:egamma2}
For $f(t) =  (\gamma - t)^+ + 1- \gamma$, we have
\begin{IEEEeqnarray}{rCl} S_r^{(f)}(D,R|{\bf X}) &=& \lim_{\nu \downarrow 0}\limsup_{n \to \infty} \frac{H_{0}(D+\nu|X^n) - nR}{\sqrt{n}},  \\
{S}_\iota^{(f)}({\Delta},R|{\bf X}) & = &  \lim_{\nu \downarrow 0}\liminf_{n \to \infty} \frac{H_{\infty}(\Delta+\nu|X^n)-nR}{\sqrt{n}}.
\end{IEEEeqnarray}
\end{corollary}
\begin{IEEEproof}
The proof is similar to proofs of Corollaries \ref{coro:vd}-\ref{coro:egamma}.
\end{IEEEproof}
The optimum achievable rates with the variational distance in terms of the smooth R\'enyi entropy have already been derived.
Relations (\ref{eq:rvd2}) and (\ref{eq:ivd2}) coincide with the result given by Tagashira and Uyematsu \cite{TU2012}, and the result given by Namekawa and Uyematsu \cite{NU2014}, respectively.
As in the case of the \textit{first-order} achievability, \textit{second-order} optimum rates for the half variational distance ($f(t) = (1-t)^+$) and the $E_\gamma$-divergence ($f(t) = (\gamma -t)^+ + 1 - \gamma$) are the same, regardless of the value of $\gamma \ge 1$.
%
%
%
%
\section{Optimistic optimum achievable rates}
\subsection{Source Resolvability}
In previous sections, we have treated general formulas of the first- and second-order optimum rates.
In this section, we consider optimum achievable rates in the optimistic sense.
The notion of the optimistic optimum rates has first been introduced by Vembu, Verd\'u and Steinberg \cite{VVS} in the source-channel coding problem.
Several researchers have developed the optimistic coding scenario in other information theoretic problems \cite{Chen1999,Hayashi,Koga2014,YN2017_7, Nomura_TIT2020}.
In this subsection, we also clarify the optimistic optimum resolvability rate with respect to the $f$-divergence using the smooth R\'enyi entropy.
\begin{definition} \label{def:7-1} 
$R$ is said to be optimistically $D$-achievable with the given $f$-divergence  if there exists a sequence of mapping $\phi_n : {\cal U}_{M_n} \to {\cal X}^n$ satisfying
\begin{equation} \label{eq:161}
\limsup_{n \to \infty}D_f(X^{n}||\phi_{n}(U_{M_{n}})) \leq D, \quad \liminf_{n \to \infty}\frac{1}{{n}} \log M_{n} \leq R.
\end{equation}
\end{definition}
\begin{definition}[Optimistic first-order optimum resolvability rate] \label{def:7-Su}
\begin{align} 
 T_r^{(f)}(D|{\bf X}) & := \inf \left\{ R \left|R \mbox{ is optimistically $D$-achievable with the given $f$-divergence}  \right. \right\}.
\end{align}
\end{definition}

We similarly define the second-order achievability in the optimistic scenario.
\begin{definition} 
{\emph L} is said to be optimistically $(D, R)$-achievable with the given $f$-divergence if there exists a sequence of mapping $\phi_n : {\cal U}_{M_n} \to {\cal X}^n$ satisfying
\begin{align}
\limsup_{n \to \infty} D_f(X^{n}||\phi_{n}(U_{M_{n}})) \leq D, \quad 
\liminf_{n \to \infty} \frac{1}{\sqrt{{n}}}\log \frac{M_{n}}{e^{{n}R}} \leq L
\end{align}
\end{definition}
\begin{definition} [Optimistic second-order optimum resolvability rate]
\begin{eqnarray} 
T_r^{(f)}(D,R|{\bf X}) := \inf \left\{ L \left|L \mbox{ is optimistically $(D, R)$-achievable with the given $f$-divergence} \right. \right\}.
\end{eqnarray}
\end{definition}
\begin{remark}
Conditions of optimistically $D$-achievability (\ref{eq:161}) can also be written as
\begin{equation} \label{eq:165}
\liminf_{n \rightarrow \infty} D_f(X^n||\phi_n(U_{M_n})) \leq D, \quad \limsup_{n \rightarrow \infty} \frac{1}{n} \log M_n \leq R.
\end{equation}
In actual, the optimistic first-order optimum resolvability rate on the basis of (\ref{eq:165}) coincides with the one defined by  Definition \ref{def:7-Su}. 
Similar argument is also applicable in the optimistic second-order optimum resolvability rate as well as the optimistic optimum intrinsic randomness rates.
\end{remark}
The following theorem can be obtained by using Theorems \ref{theo:direct} and \ref{theo:converse}.
\begin{theorem} \label{theo:7-1-1}
Under conditions C2') and C3), then for any $0 \le D < f(0)$ it holds that
\begin{align}
T_{r}^{(f)}(D|{\bf X}) & = \lim_{\nu \downarrow 0} \liminf_{n \to \infty} \frac{1}{n} H_0(1 -{f_0^{-1}}(D+\nu)|X^n),
\end{align}
{where $f_0$ is the offset function of $f$, defined in \eqref{eq:offset_function}.}
\end{theorem}
\begin{IEEEproof}
The proof proceeds in parallel with proof of Theorems \ref{theo:3-1} and \ref{theo:general3-1} in which $\limsup_{n \to \infty} 1/n \log M_n$ is replaced by $\liminf_{n \to \infty} 1/n \log M_n$.
\end{IEEEproof}

\begin{theorem} \label{theo:7-1-2}
Under conditions C2') and C3'), then for any $0 \le D < f(0)$ it holds that
\begin{align} \label{eq:7-2}
T_{r}^{(f)}(D,R|{\bf X}) & = \lim_{\nu \downarrow 0} \liminf_{n \to \infty} \frac{H_0(1-f_0^{-1}(D+\nu)|X^n) - nR}{\sqrt{n}}.
\end{align}
\end{theorem}
\begin{IEEEproof}
The proof proceeds in parallel with the proof of Theorem \ref{theo:5-1} in which 
$\limsup_{n \to \infty} 1/\sqrt{n} \log M_n$ is replaced by $\liminf_{n \to \infty} 1/\sqrt{n} \log M_n$.
\end{IEEEproof}
We have revealed the first- and second-order optimum resolvability rates in the optimistic scenario.
As a result, the effectiveness of Theorems \ref{theo:direct} and \ref{theo:converse} has also been shown.

The optimistic second-order optimum achievable rates with the half variational distance using the smooth R\'enyi entropy have already been derived by Tagashira and Uyematsu \cite{TU2012}. If we consider the case of $f(t)= (1-t)^+$, Theorem \ref{theo:7-1-2} coincides with their result.
\subsection{Intrinsic randomness}
We next consider the optimum intrinsic randomness rates in the optimistic scenario. 
\begin{definition} 
$R$ is said to be optimistically $\Delta$-achievable with the given $f$-divergence if there exists a sequence of mapping $\varphi_n : {\cal X}^n \to {\cal U}_{M_n}$ satisfying
\begin{align} 
\limsup_{n \to \infty} D_f(\varphi_{n}(X^{n})||U_{M_{n}}) & \leq \Delta, \quad  \limsup_{n \to \infty}\frac{1}{{n}} \log M_{n} \geq R.
\end{align}
\end{definition}
\begin{definition}[Optimistic first-order optimum intrinsic randomness rate] 
\begin{equation} 
T^{(f)}_{\iota}(\Delta|{\bf X}) :=  \sup \left\{ R \left|R \mbox{ is $\Delta$-achievable with the given $f$-divergence} \right. \right\}.
\end{equation}
\end{definition}
\begin{definition} 
$L$ is said to be optimistically $(\Delta,R)$-achievable with the given $f$-divergence if there exists a sequence of mapping $\varphi_n : {\cal X}^n \to {\cal U}_{M_n}$ satisfying
\begin{align} \label{eq:7-2-2-1}
\limsup_{n \to \infty} D_f(\varphi_{n}(X^{n})||U_{M_{n}}) & \leq \Delta, \quad
\limsup_{n \to \infty} \frac{1}{\sqrt{n}} \log \frac{M_{n}}{e^{{n}R}} \geq L.
\end{align}
\end{definition}
\begin{definition}[Optimistic second-order optimum intrinsic randomness rate] 
\begin{equation}
T^{(f)}_{\iota}(\Delta,R|{\bf X}) :=  \sup \left\{ L \left|L \mbox{ is optimistically $(\Delta,R)$-achievable with the given $f$-divergence} \right. \right\}.
\end{equation}
\end{definition}
Then, we have the theorem by using Theorems \ref{theo:4-1} and \ref{theo:4-2}.
\begin{theorem} \label{theo:7-2-1}
Under condition C2'), for any $0 \le \Delta < f(0)$ it holds that
\begin{align}
T^{(f)}_{\iota}(\Delta|{\bf X}) & = \lim_{\nu \downarrow 0} \limsup_{n \to \infty} \frac{1}{n} H_\infty(1-{f_0^{-1}}(\Delta+\nu)|X^n).
\end{align}
\end{theorem}
\begin{IEEEproof}
The proof is similar to the proof of Theorems \ref{theo:4-3} and \ref{theo:general4-3} in which $\liminf_{n \to \infty} 1/{n} \log M_n$ is replaced by $\limsup_{n \to \infty} 1/n \log M_n$.
\end{IEEEproof}
\begin{theorem} \label{theo:7-2-2}
Under condition C2'), for any $0 \le \Delta < f(0)$ it holds that
\begin{align}
T^{(f)}_{\iota}(\Delta,R|{\bf X}) & = \lim_{\nu \downarrow 0} \limsup_{n \to \infty} \frac{H_\infty(1-{f_0^{-1}}(\Delta+\nu)|X^n) - nR}{\sqrt{n}}.
\end{align}
\end{theorem}
\begin{IEEEproof}
The proof is similar to the proof of Theorem \ref{theo:5-2} in which $\liminf_{n \to \infty} 1/\sqrt{n} \log M_n$ is replaced by $\limsup_{n \to \infty} 1/\sqrt{n} \log M_n$.
\end{IEEEproof}
We have revealed the first- and second-order optimum intrinsic randomness rates in the optimistic scenario.
As in the case of the resolvability problem, the effectiveness of Theorems \ref{theo:4-1} and \ref{theo:4-2} has also been shown.

The optimistic first-order optimum intrinsic randomness rate with the half variational distance using the smooth R\'enyi entropy have been derived by Uyematsu and Kunimatsu \cite{Uyematsu13}, while the second-order one has been characterized by Namekawa and Uyematsu \cite{NU2014}. 
Our results (Theorems \ref{theo:7-2-1} and \ref{theo:7-2-2}) are generalizations of their results.
%
%
%
\section{Discussion}
Theorems \ref{theo:3-1} and \ref{theo:4-3} (as well as Theorems \ref{theo:5-1} and \ref{theo:5-2}) have shown a kind of \textit{duality} of two optimum achievable rates in different random number generation problems in terms of the smooth R\'enyi entropy.
It should be noted that in the case of the variational distance, Theorem 6 in \cite{Uyematsu_ISIT2010}   and Theorem 7 in \cite{Uyematsu13} have implied the same \textit{duality}.

As we have mentioned in Section I, the optimum achievable rates $S_r^{(f)}(D|{\bf X})$ and $S_\iota^{(f)}(\Delta|{\bf X})$ have already been characterized by using the information spectrum quantity.
\begin{definition} \label{def:Kf}
\begin{IEEEeqnarray}{rCl} \label{eq:Kf}
\overline{K}_{f}({\varepsilon}|{\bf X}) & := & \inf \left\{  R \left| \limsup_{n \to \infty } f\left( \Pr \left\{ \frac{1}{n} \log \frac{1}{P_{X^n}(X^n)}  \le  R  \right\} \right) \le \varepsilon  \right. \right\}, \nonumber \\
\underline{K}_f({\varepsilon}|{\bf X}) & := & \sup \left\{  R \left| \limsup_{n \to \infty } f\left( \Pr \left\{ \frac{1}{n} \log \frac{1}{P_{X^n}(X^n)} \ge  R  \right\} \right) \le \varepsilon  \right. \right\}.
\end{IEEEeqnarray}
\end{definition}
Then, using these two quantities the following theorem has already been given.
\begin{theorem}[Nomura {\cite[Theorems 3.1 and 4.1]{Nomura_TIT2020}}] \label{theo:7-1}
Under conditions C1)--C3), it holds that
\begin{IEEEeqnarray}{rCl}
S_r^{(f)}(D|{\bf X}) & = & \overline{K}_{f}(D|{\bf X}) , \\
S^{(f)}_{\iota}(\Delta|{\bf X}) &= &\underline{K}_f(\Delta|{\bf X}).
\end{IEEEeqnarray}
\end{theorem}

From the above theorem and Theorems \ref{theo:3-1} and \ref{theo:4-3}, we obtain the following relationship.
\begin{theorem}  \label{theo:7-2}
Under conditions C1)--C3), it holds that
\begin{IEEEeqnarray}{rCl}  \label{eq:7-2-1}
\lim_{\nu \downarrow 0}\limsup_{n \to \infty} \frac{1}{n}H_{0}(1 - f^{-1}(D+\nu)|X^n) & = &  \overline{K}_{f}(D|{\bf X}),   \\  \label{eq:7-2-2}
\lim_{\nu \downarrow 0}\liminf_{n \to \infty} \frac{1}{n}H_{\infty}(1 - f^{-1}(\Delta+\nu)|X^n) & = & \underline{K}_f(\Delta|{\bf X}).
\end{IEEEeqnarray}
\end{theorem}
The above theorem shows equivalences between information spectrum quantities and smooth R\'enyi entropies.
\begin{remark}  \label{remark:7-2}
Theorem \ref{theo:7-2} can also be proved by using previous results and the continuity of the function $f$.
In actual, for $f(t) = (1-t)^+$ Steinberg and Verd\'u \cite{Steinberg} have shown
\begin{equation}
S_r^{(f)}(D|{\bf X}) =   \inf \left\{  R \left| \limsup_{n \to \infty }  \Pr \left\{ \frac{1}{n} \log \frac{1}{P_{X^n}(X^n)}  \ge  R  \right\} \le D  \right. \right\},
\end{equation}
from which together with the theorem given by Uyematsu \cite[Theorem 6]{Uyematsu_ISIT2010} (Corollary \ref{coro:vd} in this paper), we obtain
\begin{equation}
 \lim_{\nu \downarrow 0}\limsup_{n \to \infty} \frac{1}{n}H_{0}(D+\nu|X^n) =   \inf \left\{  R \left| \limsup_{n \to \infty }  \Pr \left\{ \frac{1}{n} \log \frac{1}{P_{X^n}(X^n)}  \ge  R  \right\} \le D \right. \right\}.
\end{equation}
On the other hand, since 
\begin{IEEEeqnarray}{rCl}
 \overline{K}_{f}(D|{\bf X}) = 
\inf \left\{  R \left| \limsup_{n \to \infty }  \Pr \left\{ \frac{1}{n} \log \frac{1}{P_{X^n}(X^n)}  \ge  R  \right\} \le 1 - f^{-1}(D)\right. \right\}
\end{IEEEeqnarray}
holds under conditions C1)--C3), we have (\ref{eq:7-2-1}). 
Equation (\ref{eq:7-2-2}) can also be derived from Corollary \ref{coro:vd} and the result given by \cite[Theorem 2.4.2]{Han}.
\end{remark}

\begin{remark}
From Def. \ref{def:Kf}, two quantities $\overline{K}_{f}(D|{\bf X})$ and $\underline{K}_{f}(D|{\bf X})$ are right-continuous functions of $D$, while
\begin{equation}  \label{eq:166}
\limsup_{n \to \infty} \frac{1}{n}H_{0}(1 - f^{-1}(D)|X^n) \mbox{ and } \liminf_{n \to \infty} \frac{1}{n}H_{\infty}(1 - f^{-1}(D)|X^n)
\end{equation}
may not.
The operation $\lim_{\nu \downarrow 0}$ in Theorem \ref{theo:7-2} can be considered as an operation which makes quantities in (\ref{eq:166}) to be right-continuous.
Furthermore, since $f^{-1}(D)$ is a decreasing function of $D$, $H_{\alpha}(1 - f^{-1}(D)|X^n)$ is also a decreasing function of $D$.
This means that the relation
\begin{equation}
\limsup_{n \to \infty} \frac{1}{n}H_{0}(1 - f^{-1}(D)|X^n) \ge \lim_{\nu \downarrow 0} \limsup_{n \to \infty} \frac{1}{n}H_{0}(1 - f^{-1}(D+\nu)|X^n),
\end{equation}
holds. It should be emphasized that the above inequality holds with equality except for at most countably many $D$.
Similarly, we obtain
\begin{equation}
\liminf_{n \to \infty} \frac{1}{n}H_{\infty}(1 - f^{-1}(\Delta)|X^n) \ge \lim_{\nu \downarrow 0} \liminf_{n \to \infty} \frac{1}{n}H_{\infty}(1 - f^{-1}(\Delta+\nu)|X^n),
\end{equation}
where the equality holds except for at most countably many $\Delta$.
The similar observation can be applied to Theorem \ref{theo:7-4} below.
\end{remark}

We next consider the case of the \textit{second-order} setting.
We first define two quantities:
\begin{IEEEeqnarray}{rCl}  \label{eq:K_f2nd}
\overline{K}_f({\varepsilon},R|{\bf X})& :=& \inf \left\{  L \left| \limsup_{n \to \infty } f\left( \Pr \left\{ \frac{1}{n} \log \frac{1}{P_{X^n}(X^n)}  \le  R + \frac{L}{\sqrt{n}}  \right\} \right) \le \varepsilon  \right. \right\}, \\
\underline{K}_f({\varepsilon},R|{\bf X}) & :=& \sup \left\{  L \left| \limsup_{n \to \infty } f\left( \Pr \left\{ \frac{1}{n} \log \frac{1}{P_{X^n}(X^n)} \ge  R + \frac{L}{\sqrt{n}}  \right\} \right) \le \varepsilon  \right. \right\}.
\end{IEEEeqnarray}
By using these quantities, the following theorem has been obtained.
\begin{theorem}[Nomura {\cite[Theorems 6.1 and 6.2]{Nomura_TIT2020}}]
Under conditions C1), C2) and C3'), it holds that
\begin{IEEEeqnarray}{rCl}
S_r^{(f)}(D,R|{\bf X}) & = & \overline{K}_f(D,R|{\bf X}), \\
S_{\iota}^{(f)}(\Delta,R|{\bf X}) & = & \underline{K}_f(\Delta,R|{\bf X}).
\end{IEEEeqnarray}
\end{theorem}
From the above theorem and Theorems \ref{theo:5-1} and \ref{theo:5-2}, we obtain:
\begin{theorem}  \label{theo:7-4}
Under conditions C1), C2) and C3'), it holds that
\begin{IEEEeqnarray}{rCl}
\lim_{\nu \downarrow 0}\limsup_{n \to \infty} \frac{H_{0}(1 - f^{-1}(D+\nu)|X^n) - nR}{\sqrt{n}} & = & \overline{K}_f(D,R|{\bf X}), \\
\lim_{\nu \downarrow 0}\liminf_{n \to \infty} \frac{H_{\infty}(1 - f^{-1}(\Delta+\nu)|X^n)  - n R}{\sqrt{n}} & = & \underline{K}_f(\Delta,R|{\bf X}).
\end{IEEEeqnarray}
\end{theorem}
The above theorem also shows equivalences between information spectrum quantities and smooth R\'enyi entropies in the \textit{second-order} sense.

We have discussed for functions $f$ under C1), C2), and C3) (or C3')) for simplicity.
We can also extend the discussions for $f$ under C2') and C3) (or C3')) with due modification using $f_0$.
\section{Concluding Remarks}
We have so far considered the optimum achievable rates in two random number generation problems with respect to a subclass of $f$-divergences. 
We have demonstrated \textit{general formulas } of the \textit{first-} and \textit{second-order} optimum achievable rates with respect to the given $f$-divergence by using the smooth R\'enyi entropy including the inverse function of $f$.
To our knowledge, this is the first use of the smooth R\'enyi entropy in information theory that contains the general function $f$.
We believe that this is important from both of the theoretical and practical viewpoints.
In actual, we have shown that we can easily derive the results on several important measures, such as the variational distance, the KL divergence, and the Hellinger distance, by substituting the specified function $f$ into our \textit{general formulas}. 
It should be noted that the optimum achievable rates with important measures have not been characterized before by using the smooth R\'enyi entropy except the variational distance.
Expressions of the smooth max entropy in Theorem \ref{theo:Uye} and the smooth min entropy in Theorem \ref{theo:Uye2} are simple and easy to understand.
Hence, our results using the smooth max entropy and the smooth min entropy are also comprehensive.
This provides us another viewpoints to understand the mechanism of the random number generation problems compared to results given in \cite{Nomura_TIT2020}, in which the information spectrum quantities are used.
In addition, we have shown that the conditions on $f$-divergence can be relaxed, leading to the general formulas hold for a wider class of $f$-divergence. 
These are major contributions of this paper.

As a consequence of our results and results in \cite{Nomura_TIT2020}, the equivalence of the smooth R\'enyi entropy and the information spectrum quantity has been clarified (Theorem \ref{theo:7-2}).
One may consider that if we show this equivalency first, then we can derive Theorems 3.3 and 4.3 directly. 
This observation is right.
That is, one simple way to derive both of the general formulas of the optimum achievable rates (Theorems \ref{theo:3-1} and \ref{theo:4-3}) is to show this equivalency (Theorem \ref{theo:7-2}) first. Then, combining Theorem \ref{theo:7-2} and results in \cite{Nomura_TIT2020} we obtain Theorems \ref{theo:3-1} and \ref{theo:4-3}.
However, we have taken another approach to show Theorems \ref{theo:3-1} and \ref{theo:4-3} in this paper.
For example, we first have shown Theorems \ref{theo:direct} and \ref{theo:converse} so as to establish Theorem \ref{theo:3-1}.
Although Theorem \ref{theo:3-1} has been established by using Theorems \ref{theo:direct} and \ref{theo:converse}, we think that these two theorems are significant themselves.
In fact, Theorem \ref{theo:direct} provides us how to construct an optimum mapping in the resolvability problem and Theorem \ref{theo:converse} show the relationship between the rate of the random number and the smooth max entropy in terms of the finite block length.
Hence, these two theorems are also significant not only for proving Theorem 3.3 but also for constructing the optimum mapping in the practical situation.

In this paper, we have considered the $f$-divergence $D_f(X^n||\phi_n(U_{M_n}))$ in the case of the resolvability problem and $D_f(\varphi_n(X^n)||U_{M_n})$ in the case of the intrinsic randomness problem and shown a kind of \textit{duality} of these problems in terms of the smooth R\'enyi entropy.
On the other hand, we can consider the resolvability problem with respect to $D_f(\phi_n(U_{M_n})||X^n)$ 
as well as the intrinsic randomness problem with respect to $D_f(U_{M_n}||\varphi_n(X^n))$.
Although these problems are also important, the similar technique in the present paper cannot be applied directly. In order to treat these problems it seems we need some novel techniques, which remain to be studied. This is similar to the case of the information spectrum approach \cite{Nomura_TIT2020}.

Finally, the condition C3) and the assumption (\ref{eq:assump_source}) for the source, have only been needed to show Direct Part (Theorem \ref{theo:direct}) in the resolvability problem.
To consider the necessity or weaken of these conditions is also a future work.
%
%
%



%




\ifCLASSOPTIONcaptionsoff
  \newpage
\fi




\begin{thebibliography}{10}
\providecommand{\url}[1]{#1}
\csname url@samestyle\endcsname
\providecommand{\newblock}{\relax}
\providecommand{\bibinfo}[2]{#2}
\providecommand{\BIBentrySTDinterwordspacing}{\spaceskip=0pt\relax}
\providecommand{\BIBentryALTinterwordstretchfactor}{4}
\providecommand{\BIBentryALTinterwordspacing}{\spaceskip=\fontdimen2\font plus
\BIBentryALTinterwordstretchfactor\fontdimen3\font minus
  \fontdimen4\font\relax}
\providecommand{\BIBforeignlanguage}[2]{{%
\expandafter\ifx\csname l@#1\endcsname\relax
\typeout{** WARNING: IEEEtran.bst: No hyphenation pattern has been}%
\typeout{** loaded for the language `#1'. Using the pattern for}%
\typeout{** the default language instead.}%
\else
\language=\csname l@#1\endcsname
\fi
#2}}
\providecommand{\BIBdecl}{\relax}
\BIBdecl

\bibitem{NY_ISIT2020}
R.~{Nomura} and H.~{Yagi}, ``Optimum source resolvability rate with respect to
  $f$-divergences using the smooth {R}{\'e}nyi entropy,'' in \emph{Proc. 2020
  IEEE International Symposium on Information Theory (ISIT)}, 2020, pp.
  2286--2291.

\bibitem{NY_ISIT2021}
------, ``Optimum intrinsic randomness rate with respect to $f$-divergences
  using the smooth min entropy,'' in \emph{Proc. 2021 IEEE International
  Symposium on Information Theory (ISIT)}, 2021, pp. 1784--1789.

\bibitem{HV93}
T.~S. Han and S.~Verd\'u, ``Approximation theory of output statistics,''
  \emph{{IEEE} Trans. Inf. Theory}, vol.~39, no.~3, pp. 752--772, 1993.

\bibitem{Steinberg}
Y.~Steinberg and S.~Verd\'u, ``Simulation of random processes and
  rate-distortion theory,'' \emph{{IEEE} Trans. Inf. Theory}, vol.~42, no.~1,
  pp. 63--86, 1996.

\bibitem{Nomura_ISIT2018}
R.~Nomura, ``Source resolvability with {K}ullback-{L}eibler divergence,'' in
  \emph{Proc. 2018 IEEE International Symposium on Information Theory}, 2018,
  pp. 2042--2046.

\bibitem{Nomura_TIT2020}
R.~{Nomura}, ``Source resolvability and intrinsic randomness: two random number
  generation problems with respect to a subclass of $f$-divergences,''
  \emph{{IEEE} Trans. Inf. Theory}, vol.~66, no.~12, pp. 7588--7601, 2020.

\bibitem{NH2011}
R.~Nomura and T.~S. Han, ``Second-order resolvability, intrinsic randomness,
  and fixed-length source coding for mixed sources: Information spectrum
  approach,'' \emph{{IEEE} Trans. Inf. Theory}, vol.~59, no.~1, pp. 1--16,
  2013.

\bibitem{Uyematsu_ISIT2010}
T.~{Uyematsu}, ``Relating source coding and resolvability: A direct approach,''
  in \emph{Proc. 2010 IEEE International Symposium on Information Theory}, June
  2010, pp. 1350--1354.

\bibitem{VV}
S.~Vembu and S.~Verd\'{u}, ``Generating random bits from an arbitrary source:
  Fundamental limits,'' \emph{{IEEE} Trans. Inf. Theory}, vol.~41, no.~5, pp.
  1322--1332, 1995.

\bibitem{Han}
T.~S. Han, \emph{Information-Spectrum Methods in Information Theory}.\hskip 1em
  plus 0.5em minus 0.4em\relax Springer, New York, 2003.

\bibitem{Hayashi}
M.~Hayashi, ``Second-order asymptotics in fixed-length source coding and
  intrinsic randomness,'' \emph{{IEEE} Trans. Inf. Theory}, vol.~54, no.~10,
  pp. 4619--4637, 2008.

\bibitem{Uyematsu13}
T.~Uyematsu and S.~Kunimatsu, ``A new unified method for intrinsic randomness
  problems of general sources,'' in \emph{Proc. 2013 IEEE Information Theory
  Workshop (ITW)}, Sept 2013, pp. 1--5.

\bibitem{LCV2017}
J.~Liu, P.~Cuff, and S.~Verd\'u, ``${E}_\gamma$-resolvability,'' \emph{{IEEE}
  Trans. Inf. Theory}, vol.~63, no.~5, pp. 2629--2658, 2017.

\bibitem{YH_arxiv2018}
\BIBentryALTinterwordspacing
H.~Yagi and T.~S. Han, ``Variable-length resolvability for mixed sources and
  its application to variable-length source coding,'' \emph{CoRR}, vol.
  abs/1801.04439, 2018. [Online]. Available:
  \url{http://arxiv.org/abs/1801.04439}
\BIBentrySTDinterwordspacing

\bibitem{Kumagai2017a}
W.~Kumagai and M.~Hayashi, ``Second-order asymptotics of conversions of
  distributions and entangled states based on rayleigh-normal probability
  distributions,'' \emph{{IEEE} Trans. Inf. Theory}, vol.~63, no.~3, pp.
  1829--1857, 2017.

\bibitem{Kumagai2017b}
------, ``Random number conversion and {LOCC} conversion via restricted
  storage,'' \emph{{IEEE} Trans. Inf. Theory}, vol.~63, no.~4, pp. 2504--2532,
  2017.

\bibitem{YT18}
L.~{Yu} and V.~Y.~F. {Tan}, ``Simulation of random variables under {R}\'enyi
  divergence measures of all orders,'' \emph{{IEEE} Trans. Inf. Theory},
  vol.~65, no.~6, pp. 3349--3383, June 2019.

\bibitem{YH2023}
H.~Yagi and T.~S. Han, ``Variable-length resolvability for general sources and
  channels,'' \emph{Entropy}, vol.~25, no.~10, 2023.

\bibitem{csiszar2004information}
I.~Csisz{\'a}r and P.~C. Shields, ``Information theory and statistics: A
  tutorial,'' \emph{Foundations and Trends{\textregistered} in Communications
  and Information Theory}, vol.~1, no.~4, pp. 417--528, 2004.

\bibitem{SV2016}
I.~Sason and S.~Verd\'{u}, ``$f$-divergence inequalities,'' \emph{{IEEE} Trans.
  Inf. Theory}, vol.~62, no.~11, pp. 5973--6006, 2016.

\bibitem{RW2004}
R.~{Renner} and S.~{Wolf}, ``Smooth {R}\'enyi entropy and applications,'' in
  \emph{Proc. 2004 IEEE International Symposium onInformation Theory (ISIT)},
  2004, p. 233.

\bibitem{HR2011}
T.~Holenstein and R.~Renner, ``On the randomness of independent experiments,''
  \emph{{IEEE} Trans. Inf. Theory}, vol.~57, no.~4, pp. 1865--1871, 2011.

\bibitem{Uyematsu2010}
T.~Uyematsu, ``A new unified method for fixed-length source coding problems of
  general sources,'' \emph{IEICE Trans. Fundamentals}, vol. E93-A, no.~11, pp.
  1868--1877, 2010.

\bibitem{Rockafellar}
R.~T. Rockafellar, \emph{Convex Analysis}.\hskip 1em plus 0.5em minus
  0.4em\relax Princeton University Press, Princeton, 1970.

\bibitem{Hayashi2}
M.~Hayashi, ``Information spectrum approach to second-order coding rate in
  channel coding,'' \emph{{IEEE} Trans. Inf. Theory}, vol.~55, no.~11, pp.
  4947--4966, 2009.

\bibitem{Poly2010}
Y.~Polyanskiy, H.~Poor, and S.~Verd\'u, ``Channel coding rate in the finite
  blocklength regime,'' \emph{{IEEE} Trans. Inf. Theory}, vol.~56, no.~5, pp.
  2307--2359, 2010.

\bibitem{Ingber2011}
A.~Ingber and Y.~Kochman, ``The dispersion of lossy source coding,'' in
  \emph{Proc. Data Compression Conference (DCC)}, 2011, pp. 53--62.

\bibitem{KV2012}
V.~Kostina and S.~Verd\'{u}, ``Fixed-length lossy compression in the finite
  blocklength regime,'' \emph{{IEEE} Trans. Inf. Theory}, vol.~58, no.~6, pp.
  3309--3338, 2012.

\bibitem{KV2013}
I.~Kontoyiannis and S.~Verd\'{u}, ``Optimal lossless compression: Source
  varentropy and despersion,'' in \emph{Proc. 2013 IEEE International Symposium
  on Information Theory}, 2013, pp. 1739--1742.

\bibitem{TK2014}
V.~Y.~F. Tan and O.~Kosut, ``On the dispersions of three network information
  theory problems,'' \emph{{IEEE} Trans. Inf. Theory}, vol.~60, no.~2, pp.
  881--903, 2014.

\bibitem{NH2014}
R.~Nomura and T.~S. Han, ``Second-order {S}lepian-{W}olf coding theorems for
  non-mixed and mixed sources,'' \emph{{IEEE} Trans. Inf. Theory}, vol.~60,
  no.~9, pp. 5553--5572, 2014.

\bibitem{YHN2016}
H.~Yagi, T.~S. Han, and R.~Nomura, ``First- and second-order coding theorems
  for mixed memoryless channels with general mixture,'' \emph{{IEEE} Trans.
  Inf. Theory}, vol.~62, no.~8, pp. 4395--4412, 2016.

\bibitem{W2017}
S.~Watanabe, ``Second-order region for {G}ray-{W}yner network,'' \emph{{IEEE}
  Trans. Inf. Theory}, vol.~63, no.~2, pp. 1006--1018, 2017.

\bibitem{TU2012}
S.~Tagashira and T.~Uyematsu, ``The second order asymptotic rates in
  fixed-length coding and resolvability problem in terms of smooth r\'enyi
  entropy (in japanese),'' in \emph{IEICE Technical Report, IT2012-60}, 2013,
  pp. 65--70.

\bibitem{NU2014}
E.~Namekawa and T.~Uyematsu, ``The second order asymptotic rates in intrinsic
  randomness problem in terms of smooth r\'enyi entropy (in japanese),'' in
  \emph{IEICE Technical Report, IT2014-54}, 2015, pp. 1--6.

\bibitem{VVS}
S.~Vembu, S.~Verd{\'u}, and Y.~Steinberg, ``The source-channel separation
  theorem revisited,'' \emph{{IEEE} Trans. Inf. Theory}, vol.~41, no.~1, pp.
  44--54, 1995.

\bibitem{Chen1999}
P.~O. Chen and F.~Alajaji, ``{Optimistic Shannon coding theorems for arbitrary
  single-user systems},'' \emph{{IEEE} Trans. Inf. Theory}, vol.~45, no.~7, pp.
  2623--2629, 1999.

\bibitem{Koga2014}
H.~Koga, ``Four limits in probability and their roles in source coding,''
  \emph{IEICE Trans. Fundamentals}, vol.~94, no.~11, pp. 2073--2082, 2011.

\bibitem{YN2017_7}
H.~Yagi and R.~Nomura, ``Variable-length coding with cost allowing
  non-vanishing error probability,'' \emph{IEICE Trans. Fundamentals}, vol.
  E100-A, pp. 1683--1692, 2017.

\end{thebibliography}
%

\end{document}